\documentclass[letterpaper]{article}   

\usepackage[T1]{fontenc}                
\usepackage[utf8]{inputenc}             
\usepackage{lmodern}                  

\usepackage[margin=1in]{geometry}      
\usepackage{setspace}                   

\usepackage{authblk}
\author[1,2]{Valentina Barreca}
\author[2]{Giorgio Sangiovanni}
\author[1,3]{Giovanni Cantele}
\author[2]{Ludovica Zullo}
\affil[1]{Dipartimento di Fisica ``E. Pancini'', Universit\`a degli Studi di Napoli ``Federico II'', Complesso Universitario M. S. Angelo, via Cinthia 21, 80126, Napoli, Italy}

\affil[2]{Institut f\"ur Theoretische Physik und Astrophysik and W\"urzburg-Dresden Cluster of Excellence ctd.qmat,
Universit\"at W\"urzburg, 97074 W\"urzburg, Germany}

\affil[3]{CNR-SPIN, c/o University of Napoli ``Federico II'', Department of Physics, Complesso di Monte S. Angelo, via Cinthia - 80126 - Napoli, Italy}

\usepackage{amsmath,amssymb}           
\usepackage{bm}                       
\usepackage{siunitx}                    
\usepackage[version=4]{mhchem}         

\usepackage{graphicx}                   
\graphicspath{{figures/}{SI/figures/}}  
\usepackage{booktabs}                   
\usepackage[font=small,labelfont=bf]{caption}  

\usepackage[doi=true]{biblatex}   
\usepackage{xcolor}
\usepackage{hyperref}
\hypersetup{colorlinks=true, linkcolor=blue, citecolor=blue, urlcolor=blue}

\newcommand*\papertitle{Janus Asymmetry Unlocks Rashba Spintronics and Chiral Phononics in Two-Dimensional Carbo-Chalcogenides}  
\title{\papertitle}

\date{}                                 

\begin{document}

\maketitle

\begin{abstract}
Spintronics and chiral phononics offer complementary routes for nanoscale control, yet realizing both functionalities within a common two-dimensional platform remains challenging.
Here, using extensive first-principles calculations, we show that Janus asymmetry simultaneously drives both effects in transition-metal carbo-chalcogenides (JTMCCs) along two chemically distinct routes, through either chalcogen or transition-metal substitution.
Remarkably, both Janus routes yield giant Rashba responses, with $\alpha_\mathrm{R}\approx2$ eV {\AA} in Ta$_2$CSSe and NbTaCS$_2$, exceeding those reported for Mo/W-based Janus transition-metal dichalcogenides. 
Furthermore, Janus asymmetry lifts the degeneracy of opposite-chirality phonons, producing splittings as large as $\approx28$ cm$^{-1}$. More strikingly, asymmetry generates chiral carbon vibrations that were achiral in the centrosymmetric parent compounds, with splittings reaching $\approx8$ cm$^{-1}$. 
Finally, we show external electric fields tunability over Rashba strength and chiral-phonon splittings, providing a common control knob for electronic spin and lattice angular momentum. Our work offers a unified strategy for realizing spintronic and chiral phononic in two-dimensional materials.
\end{abstract}

\noindent\textbf{Keywords:} Janus monolayers; Rashba effect; chiral phonons;
carbo-chalcogenides; first-principles calculations

\section{Introduction}
Two-dimensional (2D) materials constantly draw attention due to their peculiar properties arising from reduced dimensionality and enhanced symmetry breaking \cite{Kumbhakar_2023, Bihlmayer_2022, Bordoloi_2024}. Among them, spintronics leverages the spin degree of freedom of electrons to encode, transport, and manipulate information \cite{Wolf2001}. The Rashba effect \cite{Manchon_2015} arises when spin-orbit coupling (SOC) acts in systems lacking inversion symmetry: the asymmetric crystal potential lifts the degeneracy of spin-up and spin-down bands, inducing a characteristic spin-dependent energy splitting that is linear in momentum.
Janus monolayers represent a powerful design strategy within this landscape: inversion symmetry breaking (ISB) is induced by the generated built-in out-of-plane electric field by asymmetrically decorating opposite surfaces of a 2D material with different chemical species. This intrinsic asymmetry has been shown to enhance SOC phenomena, ferroelectric polarization, piezoelectric response, and access to magnetic phases \cite{Ma2022, Zhang2022}. Yet its potential to simultaneously activate orthogonal quantum phenomena, particularly those involving the lattice degrees of freedom, remains largely unexplored.

Simultaneously, chiral phonons, lattice vibrations carrying angular momentum through circular ionic motion, have emerged as carriers of information complementary to electrons \cite{Zhang_2015,Zhu_2018,Juraschek_2025, Sun_2026}. Recent work has shown that Janus monolayers provide an intrinsic platform for ISB and chirality-related phonon phenomena \cite{Wang2026}, where the exchange of upper and lower atomic layers produces distinct chiral responses. Theoretical studies demonstrate that chiral phonons can mediate spin dynamics and generate thermal magnetization, suggesting a path beyond electron-centric spintronics \cite{Hamada2018}.

However, engineering materials where the same symmetry-breaking mechanism simultaneously and substantially enhances both effects and makes them electrically tunable, has not been demonstrated.
Here we show, by extensive first-principles calculations, how Janus asymmetry provides a unified mechanism where the intrinsic ISB simultaneously drives spintronics and chiral phononics. We identify transition metal carbo-chalcogenides (TMCCs) as our model platform. 
TMCCs are an emerging family of two-dimensional materials combining features of transition-metal dichalcogenides (TMDs) and MXenes \cite{Kassa_2024, Majed_2022, Ren_2025, Helmer_2024}, and have already been synthesized down to the monolayer limit \cite{Wang_2025, Wang_2026, Ren_2025}. Beyond their established potential for chemical and energy applications \cite{Majed_2022, Pang_2020, Jing_2019, Unnati_2024, Lakshmy_2024}, their \ce{M2CX2} stoichiometry (M=transition metal, X=chalcogen) provides an especially versatile route to Janus engineering. In contrast to conventional \ce{MX2} TMDs, where the Janus degree of freedom is naturally introduced through the chalcogen sublattice (\ce{X}), TMCCs allow inversion symmetry to be broken through either chalcogen substitution, yielding \ce{M2CXY}, or transition-metal (\ce{M}) substitution, yielding \ce{MNCX2}. This additional chemical degree of freedom considerably enlarges the design space for controlling electronic and vibrational properties through Janus asymmetry. Despite this versatility, the potential of TMCCs for spintronics and chiral phononics remains largely unexplored.
\begin{figure}
    \centering
    \includegraphics[scale=0.5]{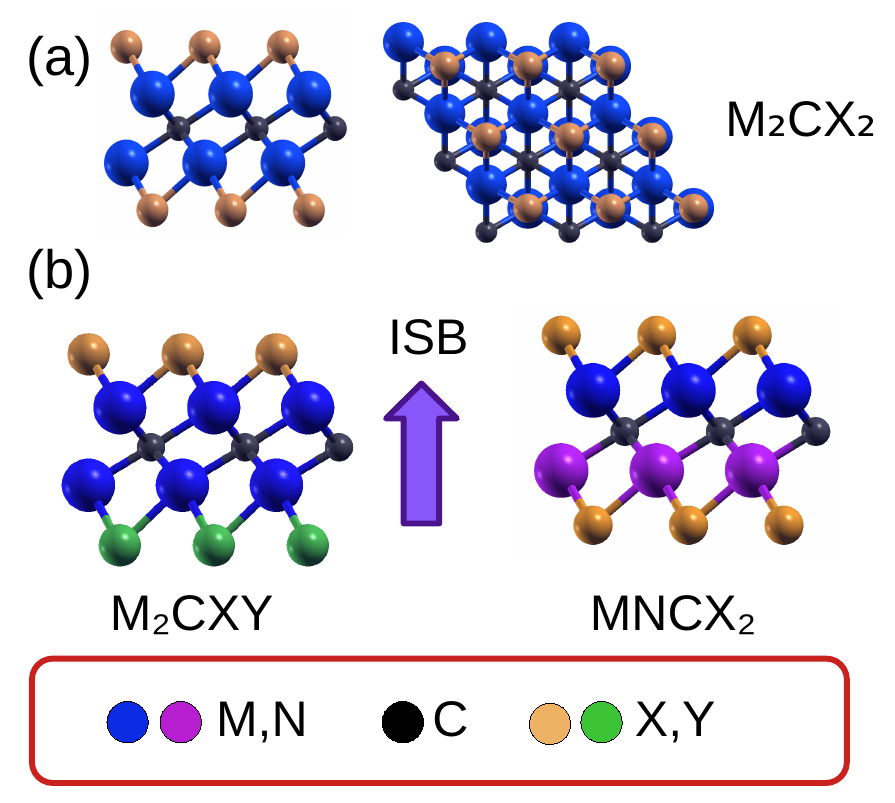}
    \caption{Structure of transition metal carbo-chalcogendides (TMCCs) and their Janus monolayers (JTMCCs). (a) Side and top view of TMCC monolayers. (b) Side view of JTMCCs monolayers. The two different atomic arrangements for Janus structure are displayed. Arrows indicate the direction of the inversion symmetry breaking (ISB) induced by atom substitution. M,N stand for transition metal atoms, X,Y for chalcogens, C for carbon. Figures made using the software XCrySDen \cite{kokalj_2003}.}
    \label{fig:fig1a}
\end{figure}

Here, we demonstrate that spintronic and chiral-phononic responses are suppressed in centrosymmetric TMCCs monolayers by symmetry protection. Even the application of strong external electric fields induces only weak Rashba effect and small chiral phonon splittings, which makes their experimental discrimination challenging. 
We show how to unlock spintronics and chiral phononics by engineering Janus-TMCCs monolayers (JTMCCs). 
Remarkably, we show giant Rashba response ($\alpha_R$ $\approx2$ eV $\cdot$ \AA) for \ce{Ta2CSSe} and \ce{NbTaCS2}, exceeding values reported for Mo/W-based Janus transition metal dichalcogenides. Crucially, we show that not only does the same inversion-symmetry breaking cause the splitting of previously degenerate chiral phonons (as large as $\approx28$ cm$^{-1}$); it also induces vibrations of carbon atoms, that were achiral in the non-Janus centrosymmetric parent compound, into right- and left-handed phonons, resulting in sizable splittings of approximately 8 cm$^{-1}$. Finally, we show how to use external electric fields to tune Rashba strength and phonon chirality in tandem, offering a single electrical knob for spin and lattice angular momentum. Our research presents a cohesive approach to achieving spintronic and chiral phononic functionalities in two-dimensional materials.

\section{Results and discussion}
The chemical formula of transition metal carbo-chalcogenides is \ce{M2CX2}, where M is the transition metal, X$=$S, Se, or Te is the chalcogen, and C is the carbon atom.
Like their parent TMDs compounds, their properties are influenced by the stacking geometry (polytypes).The H-polytype has been proven to be the preferred arrangements for metallic monolayers \cite{Helmer_2024}. As illustrated in Fig. \ref{fig:fig1a} (a), in this polytype each topmost (bottommost) X atom sits on top of the bottommost (topmost) M atom.
To explore the potential of TMCCs for 2D electronics, we focus on metallic monolayers with M$=$Nb and Ta. 
\begin{figure*}
    \centering
\includegraphics[width=0.7\linewidth]{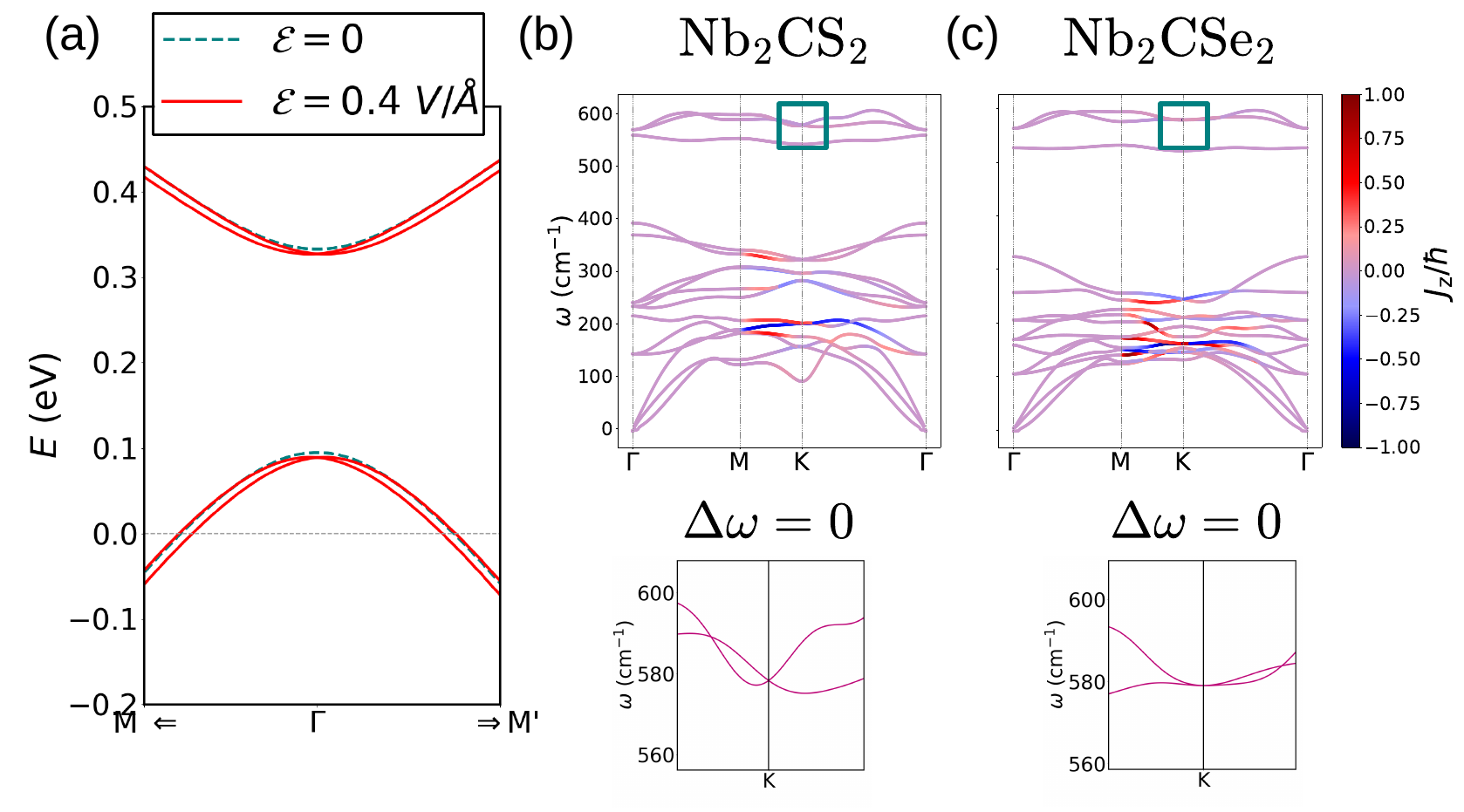}
    \caption{(a) Band structure of \protect{\ce{Ta2CSe2}} monolayer without (green dashed line) and with (red solid line) an applied electric field; zero energy is set at the Fermi level for both systems. (b)-(c) Phonon dispersions for two representative TMCC monolayers, Nb$_2$CS$_2$ and Nb$_2$CSe$_2$, respectively, colored according to the normalized phonon angular momentum $J_z/\hbar$. Blue (red) corresponds to negative (positive) circular polarization. A zoom around the $K$ point of the last two optical modes is shown. }
    \label{fig:FIG3}
\end{figure*}
To assess their application in spintronics and phononics, we start by analyzing the electronic and vibrational properties of TMCCs. First, we calculate the band structure for the family as a function of the chemistry of the X atom, using the parameters collected in Tab. S1 of SI \cite{Barreca}. The results are shown in the Fig. S2  of the SI \cite{Barreca}. As the monolayers are centrosymmetric, the spin-momentum locking is quenched by symmetry. Therefore, these TMCCs do not exhibit any Rashba spin-splittings, although potentially retaining strong SOC. 
The combination of inversion ($\mathcal{P}$) and time ($\mathcal{T}$) reversal symmetries ($\mathcal{PT}$) leads to the Kramers' spin degeneracy $E(\mathbf{k},\uparrow) =  E(\mathbf{k},\downarrow)$.  

To picture this mechanism, a representative band structure of the \ce{Ta2CSe2} compound is shown in green line of Fig. \ref{fig:FIG3}, panel (a). We consider the two sets of bands which are closer in energy to the Fermi level (from now on, $1-2$ and $3-4$). 
Even when a strong external electric field of the order of $0.4$ V/\AA \ is applied (red line in Fig. \ref{fig:FIG3}, panel (a)), the induced $\mathcal{P}$-breaking only leads to a tiny Rashba splitting. A summary of the electric-field induced Rashba splittings, measured through the Rashba coefficients $\alpha_R$ (as defined in the first section of the SI \cite{Barreca}) is given in Tab. S6 and S7 in the SI \cite{Barreca}. The electric field demonstrates limited efficacy in generating non-zero spin momentum locking, as fields around $\sim 0.3-0.4$ V/\AA \ are inadequate to produce Rashba coefficients exceeding 0.22 eV$\cdot$\AA. In practical applications, a more significant Rashba effect is desired, generated by weaker electric fields, which would minimize the risk of damaging the sample.\\
Second, we compute phonon dispersion for two representative TMCCs, namely \ce{Nb2CS2} and \ce{Nb2CSe2}, reported in Fig. \ref{fig:FIG3} (panels (b) and (c), respectively).
The calculations for the remaining considered TMCCs monolayers are reported in Fig. S10 of the SI \cite{Barreca}. The compounds as a function of X are stable and do not show any signature of charge density waves (CDWs), unlike TMDs \cite{Shen_2017, Zhang_2021, Lin_2020} where CDWs can compete with other collective phenomena.
TMCCs share the same $\mathcal{C}_3$ rotational symmetry as TMDs. To characterize the chirality of their lattice vibrations, we evaluate the phonon angular momentum associated with the ionic motion \cite{Zhang_2015, Zhu_2018, Juraschek_2025, Sun_2026}:
\begin{equation}
    \mathbf{J}_{\mathrm{ph}} = \sum_{l\alpha} m_\alpha \, \mathbf{u}_{l\alpha} \wedge \dot{\mathbf{u}}_{l\alpha},
    \label{eq:Jph}
\end{equation}
where $m_\alpha$ is the atomic mass of atom $\alpha$ and $\mathbf{u}_{l\alpha}$ is its
displacement in unit cell $l$. For each phonon mode, the out-of-plane component $J_z$ can be expressed in terms of the right- and left-circular components of the phonon eigenvector (see the SI). The normalized quantity $J_z/\hbar$ lies between $-1$ and $+1$ and measures the net circular polarization of the mode: its sign determines the handedness, while $|J_z|=\hbar$ corresponds to a fully circularly polarized vibration.
As an example, Fig. \ref{fig:FIG3} shows the phonon dispersions of \ce{Nb2CS2} and \ce{Nb2CSe2} colored according to $J_z/\hbar$ (panels (b) and (c)). Although several branches carry finite phonon angular momentum, opposite-handed partners remain degenerate at the relevant high-symmetry points in the centrosymmetric parent compounds. Since phonon angular momentum is even under spatial inversion $\mathcal{P}$ and odd under time reversal $\mathcal{T}$, the combined $\mathcal{P}\mathcal{T}$ operation relates a mode at $\mathbf{k}$ with angular momentum $J_z$ to a mode at the same $\mathbf{k}$ with $-J_z$ and the same frequency. This effect can be seen as a phonon analogue of the Kramers' spin degeneracy found in the electronic band structures. The breaking of inversion symmetry therefore removes this fixed-$\mathbf{k}$ symmetry relation and allows opposite-handed phonons to split in energy. 
Furthermore, the hybridized atomic character of chiral vibrations, illustrated in Fig. S9 of the SI \cite{Barreca}, complicates the picture.
As in the electronic case, by applying a strong external electric field ($0.4$ V/\AA), we found only small splittings of the order of $1-2$ cm$^{-1}$ between the chiral R and L partners at the $K$ point (see Tab. S8 in the SI \cite{Barreca}).
As a result, the clockwise and counterclockwise atomic circular motions are not easily spectrally resolved as separate modes, particularly without a chirality-sensitive probe, even in the presence of external fields.
\begin{figure}
    \centering
    \includegraphics[scale=0.45]{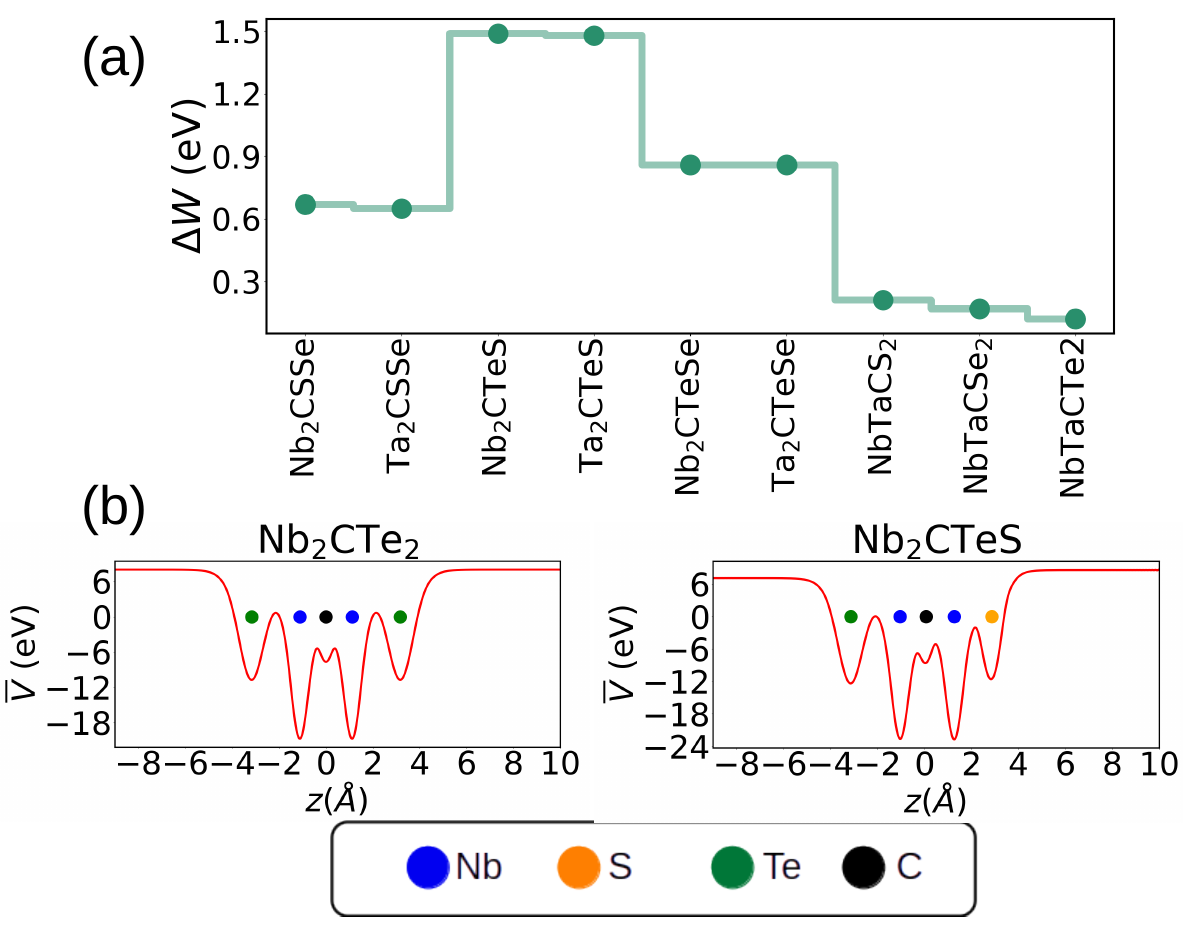}
    \caption{(a) Work function difference ($\Delta W$) between the two terminations of the JTMCCs monolayers. (b) Planar average of the electrostatic potential plotted along the out-of-plane direction for \ce{Nb2CTe2} and \ce{Nb2CTeS} monolayers. }
    \label{fig1b}
\end{figure}
We conclude that, to unlock both spintronic and chiral-phononic functionalities in TMCCs, inversion symmetry must be broken. This raises a natural question: \textit{can a single natural symmetry-breaking strategy unlock both functionalities at once?} Here, we address this question by exploiting Janus asymmetry.
Practically, Janus asymmetry is induced by substituting one atomic species with another belonging to the same group of the periodic table. For TMCCs this can be either a chalcogen or a transition-metal atom. The latter route constitutes an additional degree of freedom with respect to TMDs, where transition-metal substitution is not available within the same stoichiometric framework. Recent study reported the potential of Janus materials for storage \cite{Rivera_2026}, among which also \ce{NbTaCSe2} \cite{Rivera_2_2026}. Then, this chemical flexibility has already been explored in other contexts. Here, instead, we exploit this substitutional degree of freedom as a symmetry-engineering tool to unlock Rashba spin splitting and chiral lattice dynamics. 
\begin{figure*}
    \centering
    \includegraphics[width=0.9\linewidth]{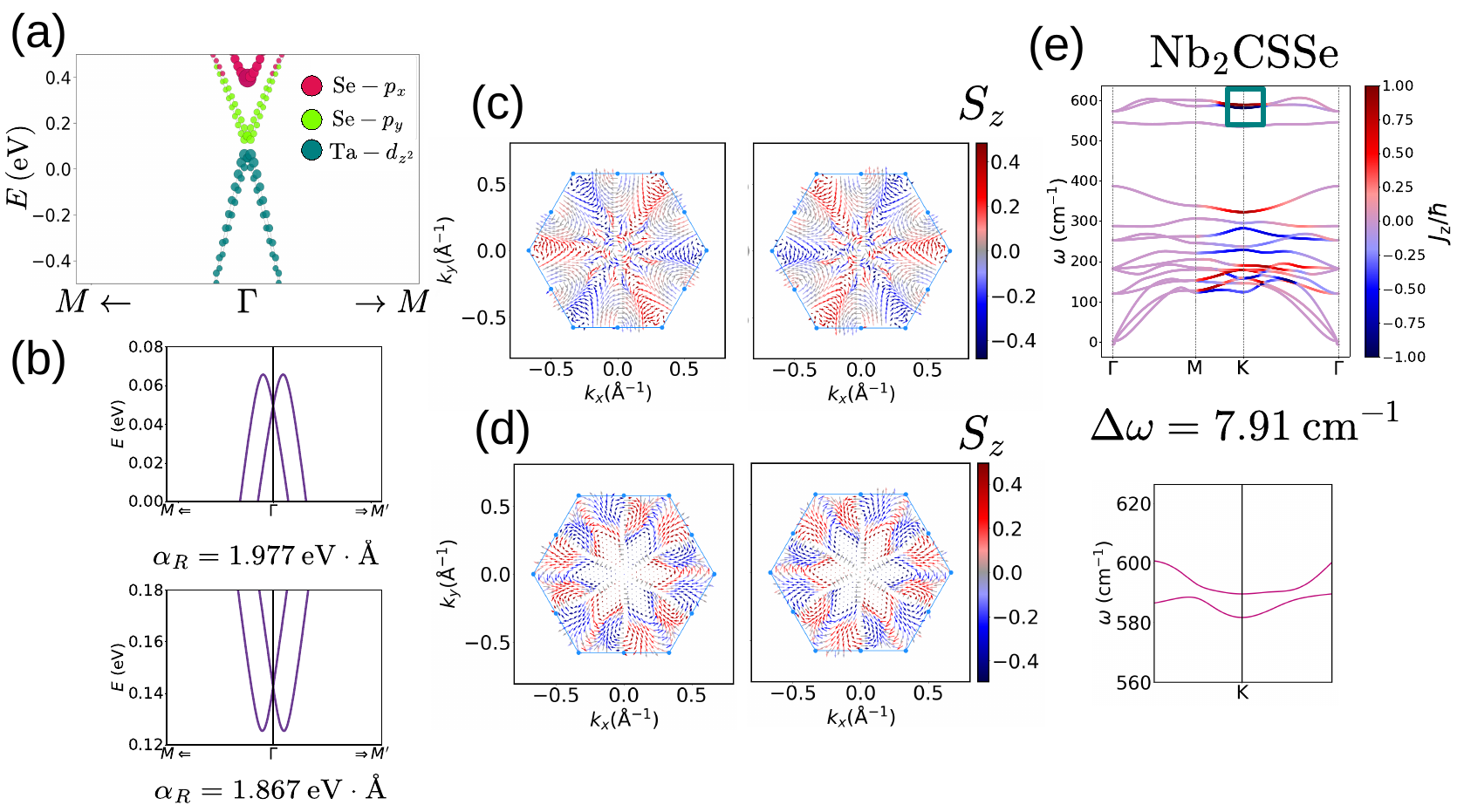}
    \caption{(a) Fatbands of \ce{Ta2CSSe} monolayer. Pink and lime dots correspond, respectively, to the $p_x$ and $p_y$ orbitals of Se, and green dots indicate $dz^2$ orbitals of Ta. The magnitude of the points is proportional to the weights of atomic orbitals on the specific wavefunction. Image made using the code QEPlotter \cite{qeplotter}. (b) Zoom of the band structure of \ce{Ta2CSSe} monolayer around $\mathrm{Ta-}dz^2$ [top] and $\mathrm{Se-}p_y$ [bottom] bands. Zero energy corresponds to Fermi level. (c) Spin-textures of $\mathrm{Ta-}dz^2$ bands of the \ce{Ta2CSSe} monolayer. (d) Spin-textures of $\mathrm{Se-}p_y$ bands of the same system. In panels (c) and (d) the colorbar corresponds to the $S_z$ component of the spin. (e) Phonon dispersion
    of \ce{Nb2CSSe} monolayer colored according to the normalized phonon angular momentum $J_z/\hbar$. Blue (red) corresponds to negative (positive) circular polarization. A zoom around the $K$ point of the last two optical modes is shown.}
    \label{fig:FIG2}
\end{figure*}
We explore all the possible Janus combinations starting from TMCCs monolayers that we have considered (JTMCCs). A sketch of the structure for a representative Janus is shown in Fig. \ref{fig:fig1a}, panel (b). Details about the choice of convergence parameters are explained in the SI \cite{Barreca}, together with the list of the optimized lattice parameters, collected in Tab. S3 of SI \cite{Barreca}.
First, we demonstrate that Janus asymmetry gives rise to a built-in asymmetric potential within the material through work function calculations. 
Fig. \ref{fig1b}, panel (b) shows the planar average of the electrostatic potential ($\bar{V}$) along the out-of-plane direction for a TMCC vs JTMCC (\ce{Nb2CTe2} vs \ce{Nb2CTeS}). It is evident that $\bar{V}$ has an asymmetric profile only for the JTMCC, indicating an internal dipole.  
We compute the work function ($W$) as in Ref. \cite{Zullo_2023}, by subtracting the Fermi energy to the value of $\bar{V}$ in the vacuum (see Tab. S10 and S11 of the SI \cite{Barreca} for the computed work functions of all the TMCC and JTMCC structures considered in this work).
TMCCs exhibit lower work functions than those of their TMD counterpart \cite{Xiaojuan_2022, Barajas_2018, Feng_2024}.
To quantitatively evaluate the Janus asymmetry, we show in Fig. \ref{fig1b}, panel (a) the work function difference ($\Delta W$) between the two out-of-plane extrema of JTMCC monolayers.
$\Delta W$ of Janus monolayers opens up the possibility for applications in  interfacial alignment for the design of heterostructures, as well as photocatalysis and directional charge transport \cite{Vu_2026, Tang_2022}.
The compounds exhibiting the smallest $\Delta W$ ($\approx 0.1–0.2$ eV) are those in which the symmetry is broken via substitution of one transition-metal atom. Instead, where a chalcogen atom is substituted, a bigger value of $\Delta W$ is obtained. The largest asymmetry is obtained for the -STe compounds, owing to the larger ionic radius mismatch between S and Te. This mismatch shortens the metal–chalcogen bond on the smaller-anion side, enhancing the structural asymmetry throughout the layer.
\begin{figure*}
    \centering
\includegraphics[width=0.8\linewidth]{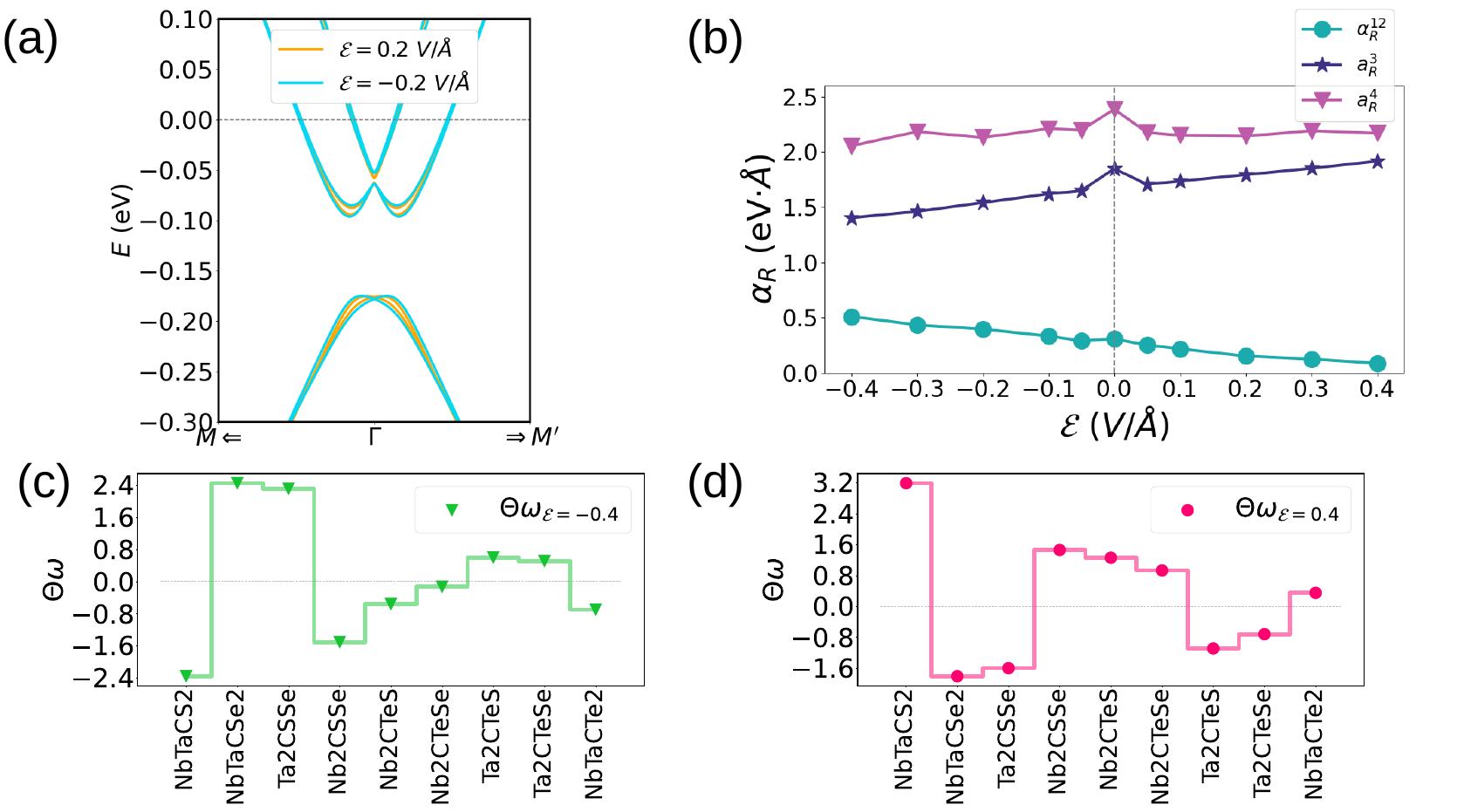}
    \caption{(a) Band structure of \ce{NbTaCS2} monolayer with the application of an external electric field along the out-of-plane direction with two different orientations. Orange and cyan lines correspond, respectively, to $\mathcal{E}=0.2$ V/\AA \ and $\mathcal{E}=-0.2 $ V/\AA \ . The Fermi level is set to zero for both systems. (b) Trend of the Rashba coefficient of \ce{NbTaCS2} monolayer as a function of the external electric field. (c-d) Variation of the splitting at $K$ point between the two highest-frequency optical chiral phonons branches in the presence of an out-of-plane electric field: $\mathcal{E}=-0.4 $ V/\AA \ (c) and $\mathcal{E}=0.4 $ V/\AA \ (d), where $\Theta\omega_{\mathcal{E}=-0.4(0.4)}=\Delta\omega_{\mathcal{E}=-0.4(0.4)}-\Delta\omega_{\mathcal{E}=0}$.  }
    \label{fig:FIG4}
\end{figure*}
Having established the effect of the intrinsic structural Janus asymmetry, we exploit its effect on electronic properties. We present their band structures in Figs. S3 and S4 of SI \cite{Barreca}. Crucially, all the Janus monolayers exhibit Rashba spin-splittings. The only exception is \ce{Nb2CTeSe}  where no bands exhibiting a sizable Rashba effect are observed around the Fermi level. Most investigated JTMCCs exhibit Janus-induced Rashba splittings in the two band pairs closest to the Fermi level, which we label $1-2$ and $3-4$ in order of increasing energy. The calculated Rashba strengths are reported in the SI (see Tables S4 and S5) \cite{Barreca}. We found that, except for \ce{Nb2CTeS}, $\alpha_{R}$ values are considerably bigger than those of the corresponding TMD Janus monolayers, being doubled or tripled, in most of the cases, with respect to the TMD counterparts \cite{Sahu_2026, Szalowski_2024}. With a few exceptions, they are also bigger than Rashba coefficients in the Janus TMDs containing Mo and W \cite{Hu_2018, Sahu_2026}. \\ 
Remarkably, \ce{Ta2CSSe} and \ce{NbTaCS2} exhibit giant Rashba effect of the order of $2$ eV $\cdot$ \AA, which is stronger than in state-of-the art TMD monolayers \cite{Yao_2017, Bordoloi_2024, Szalowski_2024}. 
A zoom around the Fermi level of \ce{Ta2CSSe} is shown in Fig. \ref{fig:FIG2}, panels (a) and (b). Giant Rashba is induced in both the last partially occupied bands and the first two unoccupied bands. 
A similar effect has been recently proved in semiconducting non-centrosymmetric TMCCs monolayers in proximity with magnetic layers \cite{Sufyan2026}.
Here, the rich chemistry of JTMCCs gives access to an emergent, atom- and orbital-resolved giant Rashba effect without the need of proximity. The fatbands in panel (a) resolve the atomic and orbital character of the Rashba-split bands. This reveals two distinct sets of spin-split states: one with dominant Ta  $d_{z^{2}}$, the other with dominant Se $p_x$/$p_y$ character. This result is significant because it implies that the atomic and orbital origin of the spin-textured carriers is not fixed, but can be selected by gating, being separated by an energy of $\approx 80\ $meV. In a 2D field-effect transistor, tuning the Fermi level allows one to populate selectively either the Ta  $d_{z^{2}}$-derived or the Se $p$-derived Rashba bands, thereby enabling an atom- and orbital-resolved switchable giant spintronic response. To derive the spin properties associated with these orbitally-distinct Rashba bands, we compute the spin textures of \ce{Ta2CSSe}, shown in Fig. \ref{fig:FIG2}, panels (c) and (d). Remarkably, the Ta $d_{z^{2}}$ and Se $\mathit{p}_x$/$p_y$ Rashba bands display different spin winding. 

This demonstrates that gate-selective population of the two band families does not merely switch which atomic species dominates transport, but corresponds to a genuine switching of the underlying spin texture. These results provide a direct signature of the atom- and orbital-resolved switchable spintronic response. Spin-textures for all the other compounds are shown in Fig. S7 and S8 of SI \cite{Barreca}.

We now examine how Janus asymmetry unlocks chiral-phonon phenomena. First, we calculate the phonon spectra of all investigated   JTMCC compounds: all monolayers result to be dynamically stable at the harmonic level, with no imaginary phonon branches. All the phonon dispersions are calculated, together with their projection onto $J_z$ (see Fig. S11 in the SI \cite{Barreca}). The dispersion of a representative compound, \ce{Nb2CSSe}, is shown in Fig. \ref{fig:FIG2}, panel (e) and can be compared with its S (Fig. \ref{fig:FIG3} panel (b)) and Se (Fig. \ref{fig:FIG3} panel (c)) TMCC counterparts.
By inspecting our calculations, we can immediately see how breaking inversion symmetry lifts the
degeneracy of opposite-handed phonons at the $K$ point. In the centrosymmetric parent compounds, the combined $\mathcal{PT}$ symmetry relates modes with opposite $J_z$ at the same wavevector and the same frequency. Once inversion symmetry is broken, only time-reversal symmetry remains: a mode at $(\mathbf{k},J_z)$ is then related to a mode at $(-\mathbf{k},-J_z)$, so opposite circular polarizations are no longer required to be degenerate at a fixed valley.
This was observed in TMDs when going from centrosymmetric bulk \ce{WSe2} to inversion-asymmetric monolayer, via infrared circular dichroism \cite{Zhu_2018} and recently predicted for Janus SbTeI \cite{Wang2026}.
For example, the splitting ($\Delta\omega$) of R and L chiral S-projected phonons at the $K$ point in the \ce{Nb2CSSe} monolayer is of the order of 28 cm$^{-1}$ (Fig. S9, panel (c) in the SI \cite{Barreca}). This is a remarkable result, since we showed that by applying an external electric field only tiny splittings are achievable.
A new impressive property comes from the last two optical C-atom projected branches. Strikingly, Janus asymmetry induces chiral phonons at the $K$ point where the TMCCs counterpart are achiral, as reported in Fig. \ref{fig:FIG2}, panel (e). This effect is general, as it appears in all the JTMCCs. We ascribe this Janus-induced chiral phonons to the internal asymmetric potential experienced by the C atoms. The C layer is the center of symmetry in the H-polytype TMCCs. Janus structure breaks its symmetry, allowing C atoms to acquire non-zero phonon angular momentum. By zooming in Fig \ref{fig:FIG2} (e), we found that R and L phonons are split at $K$. Such splittings for all the compounds are reported in the Tab. S9 of SI \cite{Barreca}. The largest splittings of Janus-induced chiral phonons appear in the \ce{Nb2CSSe} and \ce{Ta2CSSe} monolayers ($\Delta\omega \approx8$ cm$^{-1}$). 
Our results demonstrate that not only the Janus configuration splits degenerate chiral phonons, but it also induces them thanks to its intrinsic asymmetry, opening the possibility for optical detection of purely \textit{Janus-induced chiral phonons}. 

Finally, having established that Janus asymmetry activates both spintronics and chiral-phononic functionalities, we investigate how they can be further tuned by an external electric field.
Fig. \ref{fig:FIG4}, panel (a), shows the band structure around the $\Gamma$ point with two opposite electric fields ($\pm0.2$ V/\AA \ ) applied to \ce{NbTaCS2} monolayer. We note that the application of an external electric field to \ce{NbTaCS2} not only tunes the Rashba strength, but also opens a tiny gap at the $\Gamma$ point, which, strictly speaking, drives a transition between a Rashba and a mexican-hat \cite{Wick_2015,Zullo_2024_PhysRevB.110.165411} shapes of the bands. Therefore, we divide the last set of bands ($3-4$) in two ($3$ and $4$), following their mexican-hat dispersion. Unlike TMCCs, we achieve noteworthy variations of the Rashba parameter of JTMCCs (which may increase or decrease, depending on the orientation of the external electric field) with weaker fields ($\sim$ 0.1 V/\AA).
The trend for \ce{NbTaCS2} is reported in the panel (b) of Fig. \ref{fig:FIG4}.
All the results for the other JTMCCs compounds are reported in Fig. S6 in SI \cite{Barreca}.
We prove that the application of an external electric field also modifies the chiral phonons. The results obtained for the highest value of electric field ($\pm0.4$ V/\AA \ ), with two different orientations, are shown in Fig. \ref{fig:FIG4}, panels (c) and (d). Here, we report the difference in splitting of the Janus-induced carbon atoms chiral phonons between all the biased and unbiased JTMCCs, labelled as $\Theta\omega_{\mathcal{E}=-0.4(0.4)}=\Delta\omega_{\mathcal{E}=-0.4(0.4)}-\Delta\omega_{\mathcal{E}=0}$, where $\Delta\omega_{\mathcal{E}=-0.4(0.4)}$ is the splitting when we apply the external field and $\Delta\omega_{\mathcal{E}=0}$ is the splitting in the unbiased system. The sign of this variation depends on the direction of the built-in field inside the material ($-$ panel (c), $+$ panel (d)). In particular, \ce{NbTaCS2}, \ce{NbTaCSe2} and \ce{Ta2CSSe} show an increase in the splitting of the order of 2-4 $\mathrm{cm^{-1}}$. This offers another route for tuning the Janus-generated gap between R and L phonons using electric fields to facilitate their detection.

\section{Conclusions}

In conclusion, we demonstrate that Janus engineering provides a unified route to activate and control Rashba spintronics and chiral-phonon phenomena in transition-metal carbo-chalcogenides. Importantly, the structural versatility of TMCCs allows inversion symmetry to be broken through either chalcogen or transition-metal substitution, and both Janus routes yield giant Rashba responses, with $\alpha_\mathrm{R}\approx 2$ eV {\AA} in Ta$_2$CSSe and NbTaCS$_2$. In the lattice sector, Janus asymmetry lifts the degeneracy of opposite-chirality phonons, producing splittings as large as $\approx28$ cm$^{-1}$, and, more strikingly, generates chiral carbon vibrations that are achiral in the centrosymmetric parent compounds, with splittings reaching $\approx8$ cm$^{-1}$. Once these functionalities are activated by Janus asymmetry, external electric fields provide an additional means to tune both the Rashba strength and the chiral-phonon splittings. These results identify Janus carbo-chalcogenides as a chemically versatile platform in which electronic spin and lattice angular momentum can be controlled within the same two-dimensional material family. While our analysis is restricted to monolayers, the same design principles provide a natural basis for future homo- and heterostructure engineering. More generally, our findings establish
chemically encoded inversion-symmetry breaking through Janus asymmetry as a common strategy for engineering and controlling spin and chiral lattice degrees of freedom in two-dimensional materials.

\section*{Supporting Information}
Details of the pursued methodology, electronic band structure; analysis, plots and tables concerning Rashba effect and Rashba coefficient; short introduction to the phonon angular momentum, plot of phonon band structures and their projections onto atomic species, analysis and tables of the phonon frequency splittings and chiral phonons; work function discussion and analysis.

\section*{Acknowledgements}

We gratefully acknowledge financial support by the Deutsche Forschungsgemeinschaft (DFG) through the Würzburg-Dresden Cluster of Excellence ctd.qmat (EXC 2147, Project ID 390858490).
We thankfully acknowledge the CINECA HPC cluster, under the ISCRA initiative for the availability of high-performance computing resources and support. L.Z. gratefully acknowledges the scientific support and HPC resources provided by the Erlangen National High Performance Computing Center (NHR@FAU) of the Friedrich-Alexander-Universität Erlangen-N{\"u}rnberg (FAU) under the NHR project b294cb. NHR funding is provided by federal and Bavarian state authorities. The authors used generative AI tools for language editing and phrasing refinement. All scientific content, images and methodological choices are the authors' own.

\printbibliography                      

\clearpage                              

\setcounter{page}{1}
\renewcommand{\thepage}{S\arabic{page}}

\setcounter{figure}{0}
\renewcommand{\thefigure}{S\arabic{figure}}
\setcounter{table}{0}
\renewcommand{\thetable}{S\arabic{table}}
\setcounter{equation}{0}
\renewcommand{\theequation}{S\arabic{equation}}

\renewcommand{\theHfigure}{S\arabic{figure}}
\renewcommand{\theHtable}{S\arabic{table}}
\renewcommand{\theHequation}{S\arabic{equation}}
\renewcommand{\theHsection}{S\arabic{section}}

\setcounter{secnumdepth}{2}             
\setcounter{section}{0}
\renewcommand{\thesection}{\Roman{section}}

\begin{center}
  {\Large\bfseries Supporting Information for:\\[4pt] \papertitle\par}
  \vspace{6pt}
  {\normalsize Valentina Barreca, Giorgio Sangiovanni, Giovanni Cantele, Ludovica Zullo \par}  
\end{center}
\label{SI:start}                        
\vspace{12pt}

\begin{refsection}
  \section{Methods and computational details}
\begin{figure}[h]
    \centering
    \includegraphics[scale=0.6]{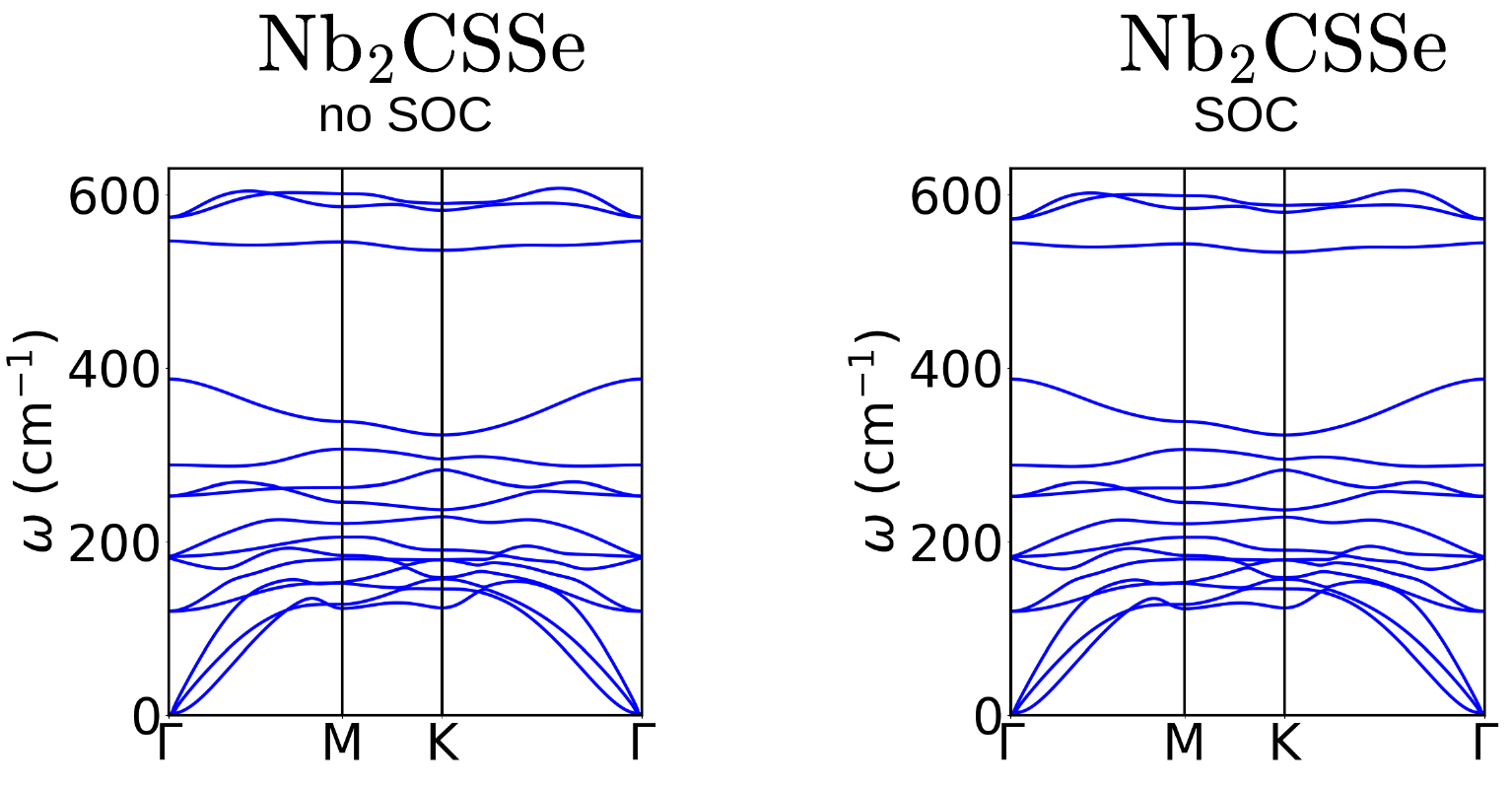}
    \caption{Comparison between the phonon dispersions obtained for the \ce{Nb2CSSe} [top]  without and with the implementation of SOC.}
    \label{soc}
\end{figure}

All simulations of structural optimizations and electronic properties were performed within the framework of Density Functional Theory (DFT), while phonon calculations by means of Density Functional Perturbation Theory (DFPT). In both cases, plane wave basis set was utilized, as implemented in the Quantum Espresso (QE) package \cite{Giannozzi_2009, Giannozzi_2017, Giannozzi_2020}.\\
Core–valence interactions were treated within the
projector augmented-wave (PAW) formalism \cite{Blochl_1994}. We employed element-specific PAW pseudopotentials \cite{Schwerdtfeger_2011} from the PSlibrary (versions $0.1$ for C and S, $0.3.0$ for Nb and $1.0.0$ for Se, Ta and Te) \cite{DalCorso_2014}, using either scalar-relativistic (optimization of atomic positions, phonons dispersions) or fully relativistic (band structures, work functions) datasets, according to the computed property. Exchange–correlation effects were described with the generalized gradient approximation in the Perdew–Burke–Ernzerhof (PBE) parametrization \cite{Ernzerhof_1999}. We chose to compute phonon dispersion without including spin-orbit coupling (SOC) as dispersions are expected to not exhibit significant dependence on SOC, while the computational cost increases notably. We verified this assumption by performing tests on selected compounds; in Fig. \ref{soc}, we present the results for \ce{Nb2CSSe} as an example.\\
Structural relaxations were performed using the Broyden–Fletcher–Goldfarb–Shanno algorithm until the total energy between successive steps changed below $10^{-4}$ Ry and all force components were smaller in modulus than $10^{-3}$ Ry/a.u. For self-consistent calculations we used a total-energy threshold of $10^{-10}$ Ry, while for the phonon calculations of $10^{-16}$ Ry.\\
To avoid interactions between periodic replicas, the supercell approach was employed, where the length of the cell in the out-of-plane direction was selected between 16 and 20 \AA, as determined through convergence tests. For asymmetric structures, a dipole correction was applied to suppress spurious electrostatic interactions between periodically repeated slabs along the out-of-plane direction \cite{Bengtsson_1999}, particularly in the calculation of work functions.
For the calculation of spin-textures we used non-collinear DFT, in which Kohn-Sham wave functions become spinors, and computed the expected value of the three components of the spin for each $k$ in the BZ.
External-field calculations were also performed with QE including SOC, using its saw-tooth potential implementation. For each different value of the electric field we relaxed the atomic positions, obtaining in all cases negligible differences with respect to the unperturbed case.\\ 

Projections of phonons on the constituent atomic species are obtained from the weights of each eigenfunction, for a specific atomic element and in all the possible directions. 
The mode-resolved phonon angular momentum along the out-of-plane direction, equivalently the phonon circular polarization $J_z/\hbar$, is computed from
the DFPT eigenvectors. A more comprehensive discussion is given in Sec. V.\\
Convergence tests were performed separately for the simulations of electronic and vibrational properties. Parameters used for the calculation of phonons allow a convergence on the frequencies smaller than 1 $\mathrm{cm^{-1}}$.\\
All convergence parameters utilized for TMCC monolayers are collected in the Tables \ref{conv_param_elec} for the electronic properties, including also the optimized lattice parameter, and in Table \ref{conv_param_ph} for the phonons branches.
\begin{table}[h]
    \centering
\begin{tabular}{ |p{2.2cm}||p{3cm}|p{2.2cm}| p{2.6cm}|p{2.2cm}| }
 \hline
     & PW cutoff (Ry) &  $n_k$ & smearing (Ry)&  $a$ (\AA) \\
 \hline
 \ce{Nb2CS2} & 50 & 16 & 0.03 &  3.28 \\
 \hline
 \ce{Nb2CSe2} & 60 & 16 & 0.03  & 3.33\\
 \hline
 \ce{Nb2CTe2}  & 60& 16 &0.03  & 3.42 \\
 \hline
 \ce{Ta2CS2}& 60 & 16 &0.035  & 3.28\\
 \hline
 \ce{Ta2CSe2}& 50  & 14 & 0.035  & 3.33\\
 \hline
  \ce{Ta2CTe2}& 60   & 16 & 0.03  & 3.43 \\
 \hline
\end{tabular}
\caption{Convergence parameters utilized for the calculation of electronic properties of TMCCs. In order: plane-waves cutoff, $n_k\times n_k \times 1$ $k$-point grid and smearing; the last column corresponds to the optimized lattice parameter.}
\label{conv_param_elec}
\end{table}

\begin{table}[h]
    \centering
\begin{tabular}{ |p{2.2cm}||p{3cm}|p{2.2cm}| p{2.6cm}|p{2.2cm}|}
 \hline
     & PW cutoff (Ry) & $n_k$ & smearing (Ry) &  $n_q$ \\  
 \hline
 \ce{Nb2CS2} & 60 & 16 & 0.03 &  12  \\
 \hline
 \ce{Nb2CSe2} & 60 & 18 & 0.035  & 12 \\
 \hline
 \ce{Nb2CTe2}  & 60 & 18 &0.025 & 18 \\
 \hline
 \ce{Ta2CS2}& 60 & 18 & 0.03  & 12 \\
 \hline
 \ce{Ta2CSe2}& 60 & 18 & 0.03  & 12  \\
 \hline
  \ce{Ta2CTe2}& 60  & 18 & 0.025 & 18  \\
 \hline
\end{tabular}
\caption{Convergence parameters utilized for the calculation of phonons frequencies of TMCCs. In order: plane-waves cutoff, $n_k\times n_k\times 1$ $k$-point grid, smearing and $n_q\times n_q \times 1$ $q$-point grid.}
\label{conv_param_ph}
\end{table}
From Table \ref{conv_param_elec} we can also see that the lattice parameter depends substantially on the chalcogen atom.\\
The convergence parameters for Janus compounds are selected in accordance to the tightest parameters for the parent TMCC compounds (for example, for \ce{Nb2CSSe} we chose between the parameters used for \ce{Nb2CS2} and \ce{Nb2CSe2}).\\
In Table \ref{tmccj_lattice} we collect the optimized lattice parameter for JTMCCs monolayers.
\begin{table}[h]
    \centering
\begin{tabular}{ |p{2.2cm}||p{2.2cm}| }
 \hline
    & a (\AA) \\
 \hline
  \ce{Nb2CSSe}   & 3.30 \\
 \hline
 \ce{Nb2CTeS}  &    3.35 \\
 \hline
 \ce{Nb2CTeSe} & 3.38 \\
 \hline
 \ce{Ta2CSSe}  & 3.30 \\
 \hline
 \ce{Ta2CTeS}& 3.35   \\
 \hline
 \ce{Ta2CTeSe}&   3.38 \\
 \hline
 \ce{NbTaCS2} & 3.28 \\
 \hline
  \ce{NbTaCSe2}&   3.33 \\
 \hline
  \ce{NbTaCTe2}& 3.42  \\
 \hline
\end{tabular}
\caption{Optimized lattice parameter for JTMCCs monolayers.}
\label{tmccj_lattice}
\end{table}
Also in this case, the dependence on the specific choice of chalcogen atoms is clear: when the symmetry is broken via the substitution of a chalcogen atom, lattice parameters are the same for the compounds with the same terminations; in the other three cases, the lattice parameter is the same as in the TMCCs with the same chalcogen atom.\\
Rashba coefficients are computed along the $M-\Gamma-M'$ direction, after the calculation of band structures around $\Gamma$-point with 200 points per segment. We identified $E_R$ as shift in energy between the minimum (or maximum, depending on the band curvature) of the energy with respect the energy at the $\Gamma$-point, while $k_R$ is the shift of the momentum in the BZ. Rashba coefficients are then computed as $\frac{2E_R}{k_R}$.\\
Spin textures are computed from the average value of spin components, as explained in Section IV.\\ 
The chemical formula of the Janus compounds are of the type \ce{M2CXY} or \ce{MNCX_2}, depending on the case, where M,N=Nb,Ta, X,Y=S,Se,Te and C is the Carbon atom. The chosen convention about the order of elements of the same types in chemical formula is based on their relative positions. Then, the first appearing is the one with the smallest out-of-plane coordinate, while the second has the highest. Then, in \ce{MNCX_2}, M has a smaller coordinate with respect to N; it is the same for X and Y in compounds of the type \ce{M2CXY}. The positive external electric field goes from the side with the element with the smallest coordinate (M, X) to the one with the highest (X,Y).\\
\clearpage
\section{Band structures}
\begin{figure}[h]
    \centering
    \includegraphics[scale=0.8]{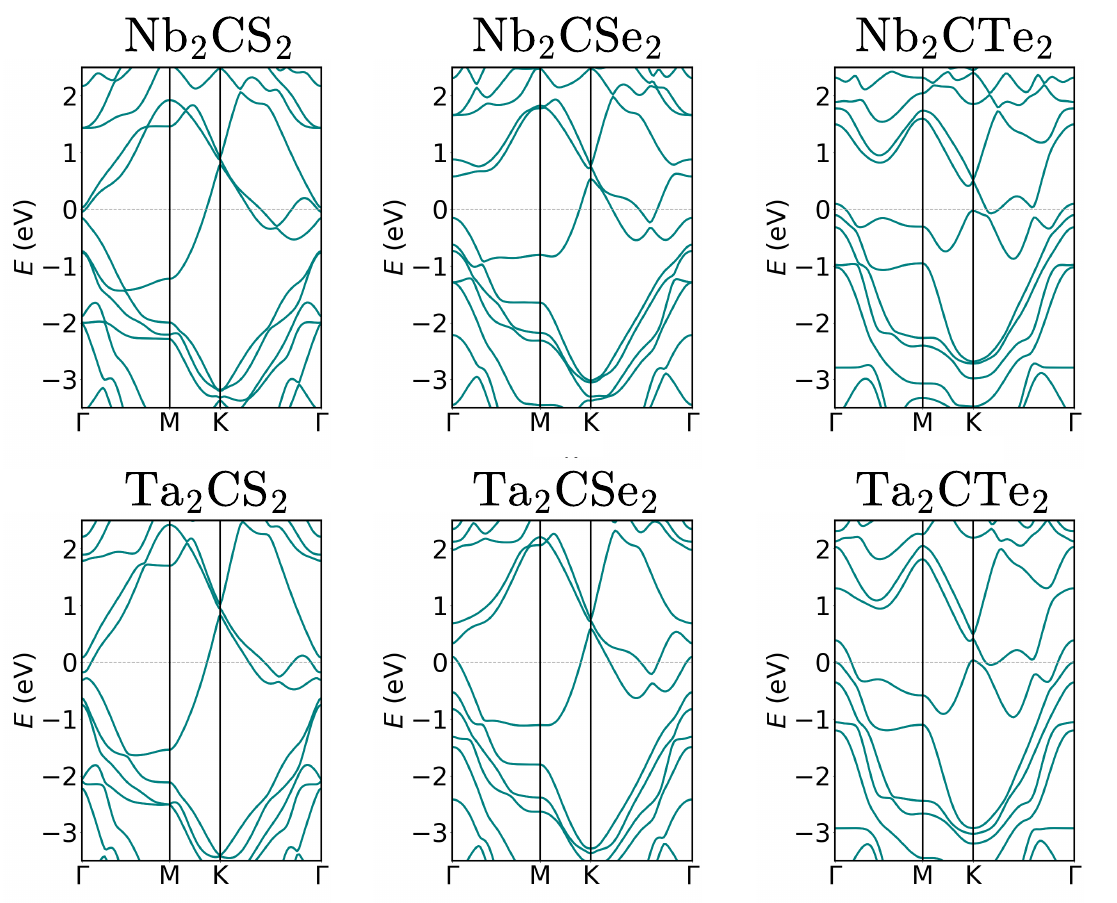}
    \caption{Band structures along the $\Gamma-M-K-\Gamma$ path for the six TMCCs monolayers analyzed in this work. SOC is included in these calculations. Zero energy corresponds to Fermi level.}
    \label{fig:tmccsbands}
\end{figure}

\begin{figure}[h]
    \centering
    \includegraphics[scale=0.8]{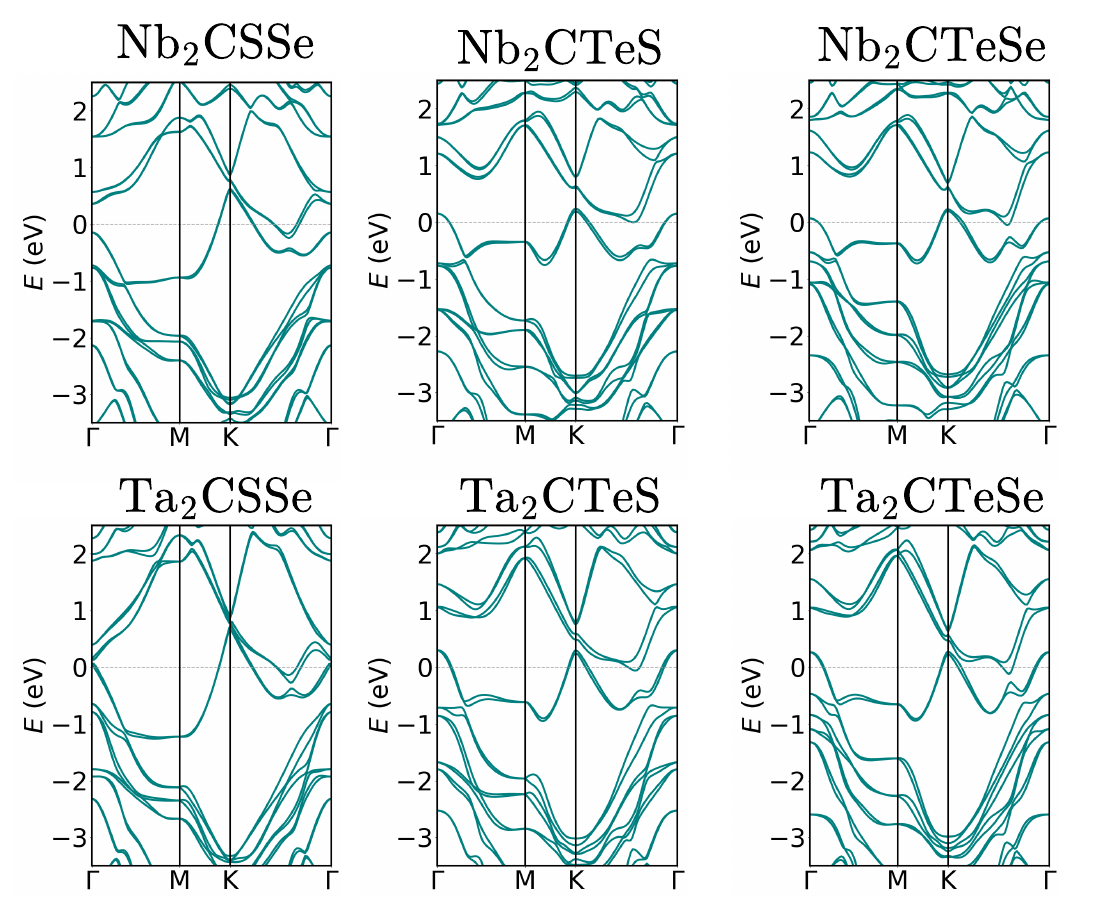}
    \caption{Band structures along the $\Gamma-M-K-\Gamma$ path for the six JTMCCs monolayers obtained breaking the inversion symmetry via the substitution of one of the chalcogen atoms. SOC is included in these calculations. Zero energy corresponds to Fermi level.}
    \label{fig:tmccj1}
\end{figure}

\begin{figure}[h]
    \centering
    \includegraphics[scale=0.8]{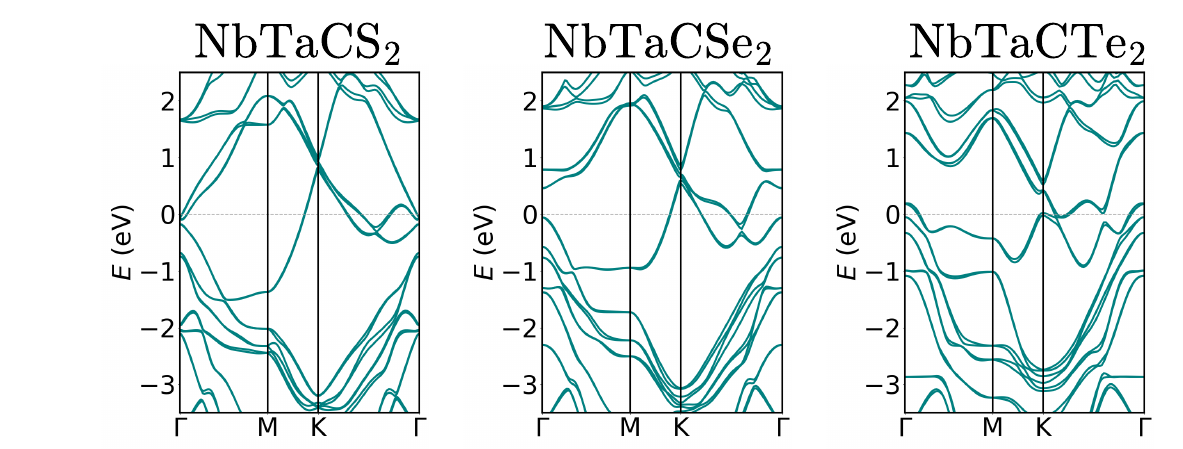}
    \caption{Band structures along the $\Gamma-M-K-\Gamma$ path for the three JTMCCs monolayers obtained breaking the inversion symmetry via the substitution of one of the transition-metal atoms. SOC is included in these calculations. Zero energy corresponds to Fermi level.}
    \label{fig:tmccj2}
\end{figure}

In this section we show the band structures for all considered monolayers along the $\Gamma-M-K-\Gamma$ path in the first BZ. In Fig. \ref{fig:tmccsbands} we report the results for TMCC monolayers. The inversion symmetry prevents SOC-induced splittings.\\
In Fig. \ref{fig:tmccj1} and \ref{fig:tmccj2} we show the band structures of the JTMCC monolayers, respectively in the cases in which the inversion symmetry is broken via the substitution of a chalcogen or a transition metal.
Splittings due to the presence of SOC are evident, especially in systems where heavy chemical species, as Te, are present. 
\clearpage
\section{Rashba effect}
\begin{figure}
    \centering
    \includegraphics[width=0.7\linewidth]{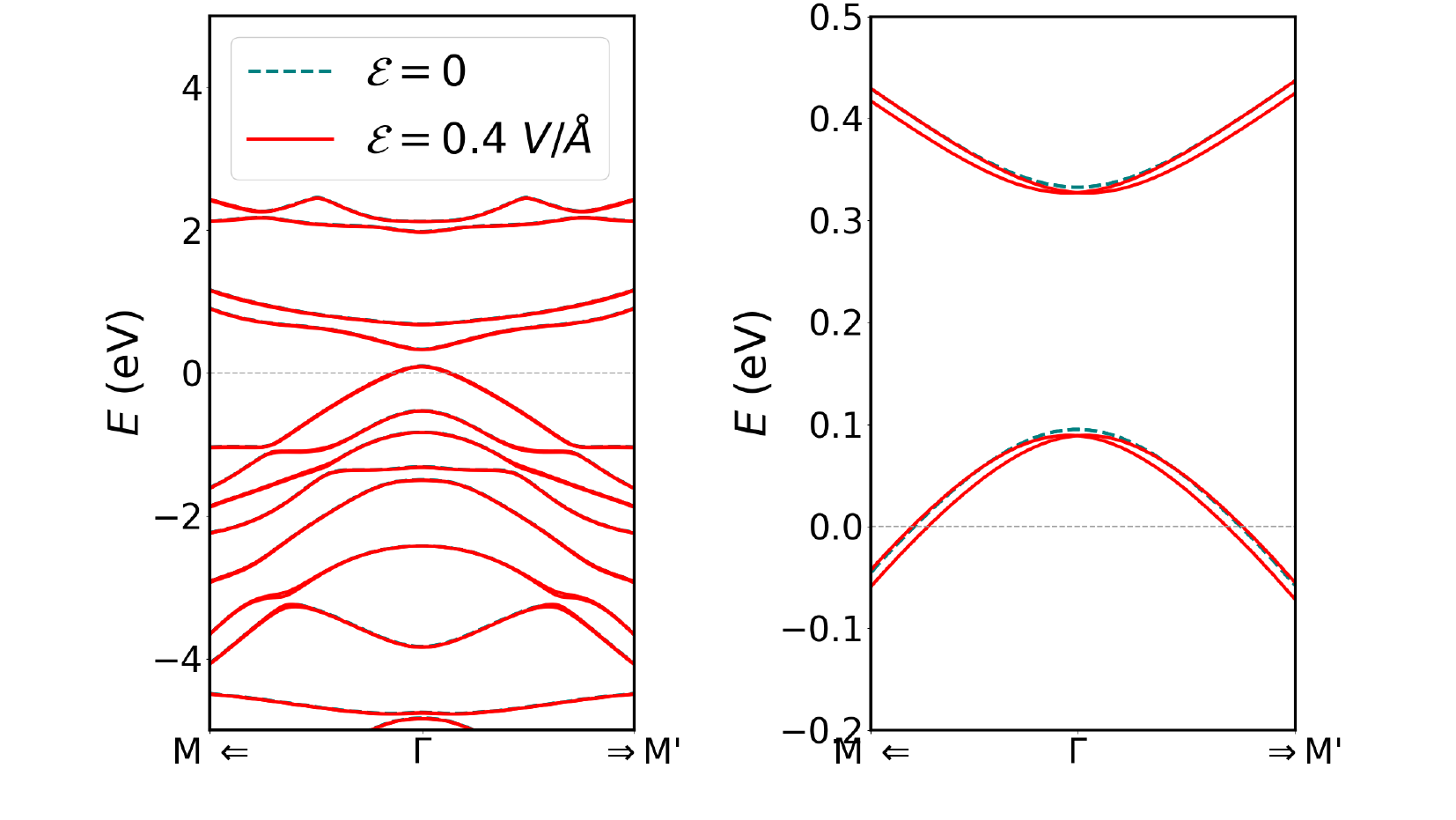}
    \caption{ Band structure of \protect{\ce{Ta2CSe2}} monolayer without (green dashed line) and with (red solid line) an applied electrostatic field. The tiny electric-field-induced splittings are visible in the zoom in the right panel. The zero of the energy is set on Fermi level of both systems.}
    \label{fig:rashba_example}
\end{figure}

\begin{figure}
    \centering
    \includegraphics[scale=0.65]{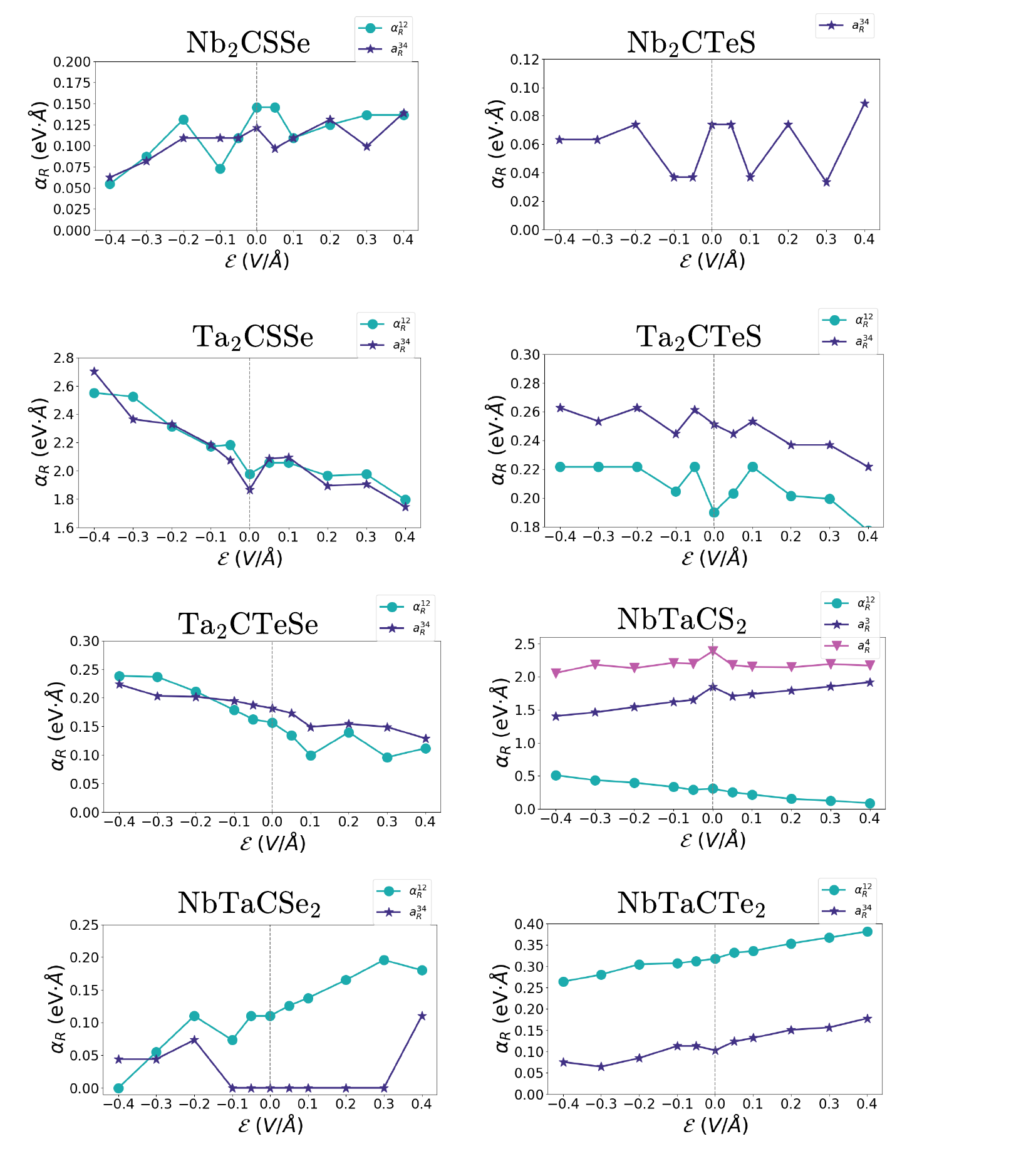}
    \caption{Rashba coefficients of the JTMCCs monolayers as a function of an external electric field. Coefficients are expressed in eV$\cdot$\AA.}
    \label{rashbavsE}
\end{figure}

The presence of splittings in the band structures of JTMCCs, together with the breaking of inversion symmetry, opens the possibility for the emergence of Rashba effect, whose coefficients are collected in Tables \ref{tmccj_rashbav} and \ref{tmccj_rashbac}, for the two pairs of bands that are the closest to the Fermi level. We indicate them as 1-2 and 3-4.
\begin{table}[h]
    \centering
\begin{tabular}{ |p{2.2cm}||p{2.2cm}|p{2.2cm}|p{2.2cm}| }
 \hline
 \multicolumn{2}{|c|}{JTMCCs (1-2)} \\
 \hline
         & $\alpha_R$ \\
\hline
  \ce{Nb2CSSe}   &  0.145 \\
 \hline
 \ce{Nb2CTeS}  &   0.000   \\
 \hline
 \ce{Nb2CTeSe} &   0.000\\
 \hline
 \ce{Ta2CSSe}  &  1.977 \\
 \hline
 \ce{Ta2CTeS}&   0.190   \\
 \hline
 \ce{Ta2CTeSe}&   0.156 \\
 \hline
 \ce{NbTaCS2} &  0.307 \\
 \hline
  \ce{NbTaCSe2}&   0.110 \\
 \hline
  \ce{NbTaCTe2}&  0.318  \\
 \hline
\end{tabular}
\caption{Rashba coefficients with $\mathcal{E}=$0 for bands 1-2 of JTMCCs monolayers. Rashba coefficients are expressed in eV$\cdot$\AA.}
\label{tmccj_rashbav}
\end{table}
\begin{table}[h]
    \centering
\begin{tabular}{ |p{2.2cm}||p{2.2cm}|p{2.2cm}|p{2.2cm}|  }
 \hline
 \multicolumn{2}{|c|}{JTMCCs (3-4)} \\
  \hline
         &  $\alpha_R$ \\
 \hline
  \ce{Nb2CSSe}   &  0.121  \\
 \hline
 \ce{Nb2CTeS}  &  0.739    \\
 \hline
 \ce{Nb2CTeSe} &  0.000\\
 \hline
 \ce{Ta2CSSe}  & 1.867 \\
 \hline
 \ce{Ta2CTeS}&   0.251  \\
 \hline
 \ce{Ta2CTeSe}&   0.182  \\
 \hline
 \ce{NbTaCS2}&  1.848/2.387 \\
 \hline
  \ce{NbTaCSe2}&  0.000  \\
 \hline
  \ce{NbTaCTe2}& 0.103  \\
 \hline
\end{tabular}
\caption{Rashba coefficients with $\mathcal{E}=$0 for bands 3-4 of JTMCCs monolayers. $E_R$ are expressed in eV.}
\label{tmccj_rashbac}
\end{table}
Coefficients of the \ce{NbTaCS2} monolayer are not reported in the previous Tables, as in this case we have three pairs of bands near to  Fermi level, showing Rashba effect. In this case Rashba coefficients are, respectively, 0.307 $\mathrm{eV}\cdot$\AA\hspace{0.5mm} for bands 1-2, $2.387\ \mathrm{eV}\cdot$\AA\hspace{0.5mm} for band 3, and $1.848\ \mathrm{eV}\cdot$\AA\hspace{0.5mm} for band 4 (see main text).\\
Two systems exhibit a giant Rashba effect: \ce{Ta2CSSe} in both pairs of bands, as described in the main text, and \ce{NbTaCS2}.\\
In Tables \ref{tmcc_rashbav} and \ref{tmcc_rashbac} the Rashba coefficients for the TMCC monolayers computed in the presence of an out-of-plane external electric field are collected; by symmetry, we explored only one orientation of the field. To induce the Rashba effect, we need strong electric fields, but the effect itself is weak. Just as an example of that, in the Fig. \ref{fig:rashba_example}, we report the band structure of \ce{Ta2CSe2} monolayer without [green line] and with [red line] an applied electric field. \\
\begin{table}[h]
    \centering
\begin{tabular}{ |p{2.2cm}||p{2.2cm}|p{2.2cm}|p{2.2cm}|p{2.2cm}|p{2.2cm}|p{2.2cm}|  }
 \hline
 \multicolumn{7}{|c|}{TMCCs (1-2)} \\
 \hline
   & $\mathcal{E}=$0.0 & $\mathcal{E}=0.05 V/\AA $& $\mathcal{E}=$0.1 V/\AA & $\mathcal{E}=$0.2 V/\AA & $\mathcal{E}=$0.3 V/\AA & $\mathcal{E}=$0.4 V/\AA \\
 \hline
 \ce{Nb2CS2}   & 0.000 & 0.000  & 0.000 & 0.000 & 0.000 & 0.000\\
 \hline
 \ce{Nb2CSe2}  & 0.000 & 0.000 & 0.000 & 0.000 & 0.073 & 0.000 \\
 \hline
 \ce{Nb2CTe2} & 0.000 & 0.000  & 0.000 & 0.000 & 0.000 & 0.000\\
 \hline
 \ce{Ta2CS2}  & 0.000 & 0.000 & 0.000 & 0.000 & 0.000 & 0.000 \\
 \hline
 \ce{Ta2CSe2}& 0.000   &  0.000 & 0.000 & 0.073& 0.147 & 0.220\\
 \hline
 \ce{Ta2CTe2}& 0.000  & 0.000 & 0.000 & 0.000 & 0.076 & 0.057 \\
 \hline
\end{tabular}
\caption{Rashba coefficients for increasing values of the intensity of the out-of-plane external field is modified for TMCC monolayers. Results obtained for one of the pairs of bands near to Fermi level, labelled as 1 and 2. Coefficients are expressed in eV$\cdot$\AA.\label{tmcc_rashbav}}
\end{table}
\begin{table}[h]
    \centering
\begin{tabular}{ |p{2.2cm}||p{2.2cm}|p{2.2cm}|p{2.2cm}|p{2.2cm}|p{2.2cm}|p{2.2cm}|  }
 \hline
 \multicolumn{7}{|c|}{TMCCs (3-4)} \\
 \hline
   & $\mathcal{E}=$0.0 & $\mathcal{E}=0.05 V/\AA $& $\mathcal{E}=$0.1 V/\AA & $\mathcal{E}=$0.2 V/\AA & $\mathcal{E}=$0.3 V/\AA & $\mathcal{E}=$0.4 V/\AA \\
 \hline
 \ce{Nb2CS2}   & 0.000 & 0.000  & 0.000 & 0.000 &0.108  & 0.000\\
 \hline
 \ce{Nb2CSe2}&  0.000   & 0.000 & 0.000 & 0.000 & 0.055 & 0.044 \\
 \hline
 \ce{Nb2CTe2} & 0.000 & 0.000  & 0.000 & 0.000 & 0.000 & 0.000\\
 \hline
 \ce{Ta2CS2}  & 0.000 & 0.000 & 0.000 & 0.000 & 0.000 & 0.000\\
 \hline
 \ce{Ta2CSe2}&  0.000  & 0.000 & 0.000 &0.110 & 0.110 & 0.132\\
 \hline
 \ce{Ta2CTe2}& 0.000  & 0.000 & 0.000 & 0.000 & 0.068 & 0.065  \\
 \hline
\end{tabular}
\caption{Rashba coefficients as the intensity of the out-of-plane external field is modified for TMCC monolayers. Results obtained for one pair of bands near to Fermi level, indicated as 3 and 4. The coefficients are expressed in eV$\cdot$\AA. }
\label{tmcc_rashbac}
\end{table}
In Fig. \ref{rashbavsE} we show the trend of Rashba coefficients as a function of external electric field for JTMCCs monolayers. \ce{Nb2CTeSe} is missing as it does not show Rashba effect until we apply a field of the order of 0.4 V/\AA. Even at such a strong electric field, the effect is still weak for this compound.
\section{Spin-textures}
\begin{figure}
    \centering
\includegraphics[scale=0.53]{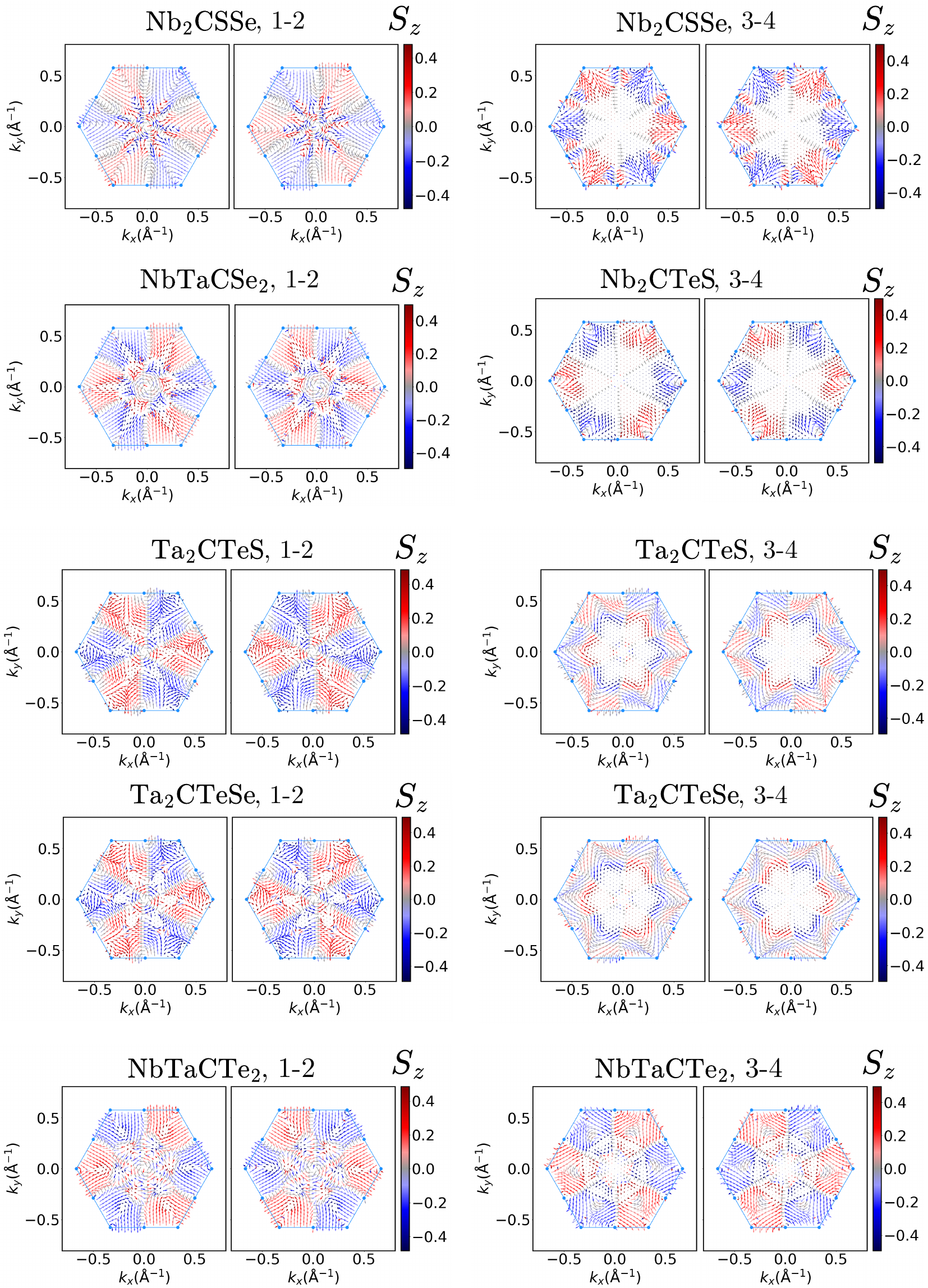}
    \caption{Spin-textures for bands 1-2 and 3-4 of all the monolayers that show Rashba effect, except \ce{Ta2CSSe}, shown in the main text, and \ce{NbTaCS2}, that is the following figure. The colorbar corresponds to the $S_z$ component of the spin}
    \label{texture1}
\end{figure}

\begin{figure}
    \centering
    \includegraphics[scale=0.53]{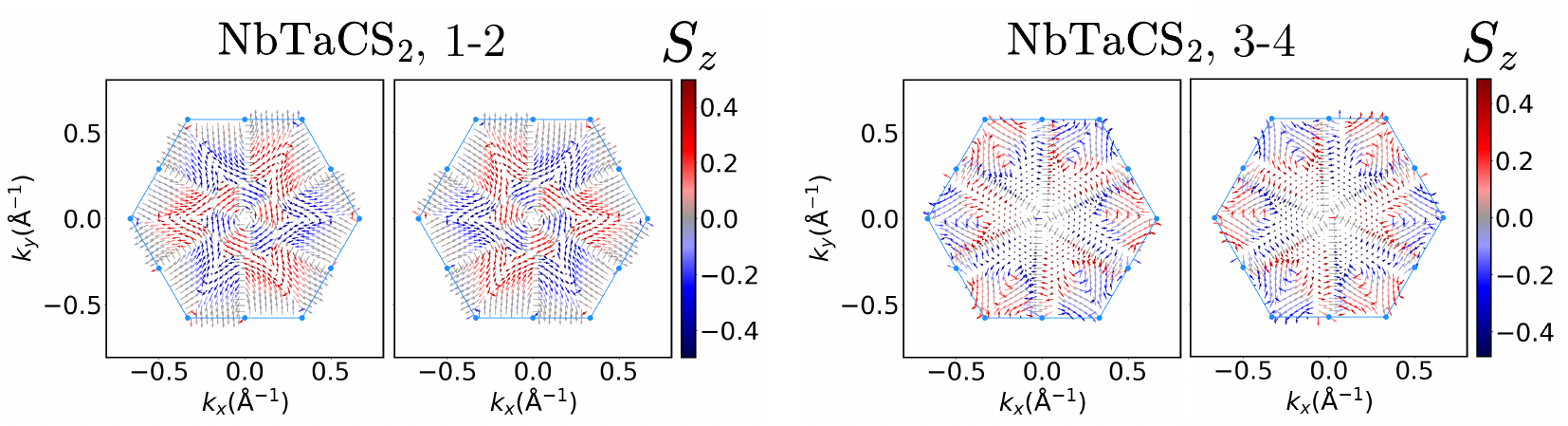}
    \caption{Spin textures for bands 1-2 and 3-4 of \ce{NbTaCS2} monolayer. The colorbar corresponds to the $S_z$ component of the spin}
    \label{texture2}.
\end{figure}

In this section we collect the spin-textures of all unbiased systems showing Rashba effect (Fig. \ref{texture1} and \ref{texture2}).
In the plots, the $x$ and $y$ components of the arrows are proportional to, respectively, the average of $S_x$ and $S_y$ components of the spin (then, their length $S_{\parallel}=\sqrt{S_x^2+S_y^2}$ is the average magnitude of the in-plane spin component).
Colorbar, instead, indicates the value of the averaged out-of-plane component $S_z$.\\

\section{Chiral Phonons}
\subsection{Phonon angular momentum and chirality}
\newcommand{\enu}{\mathbf{e}_{\alpha\nu}(\mathbf{k})}
\newcommand{\enux}{e_{\alpha\nu}^{x}(\mathbf{k})}
\newcommand{\enuy}{e_{\alpha\nu}^{y}(\mathbf{k})}
\newcommand{\epsnuR}{\epsilon_{\alpha\nu}^{R}(\mathbf{k})}
\newcommand{\epsnuL}{\epsilon_{\alpha\nu}^{L}(\mathbf{k})}
A phonon mode can carry angular momentum when the ionic motion contains an elliptical or
circular component in the plane perpendicular to the quantization axis. Importantly, non-degeneracy does not imply linear ionic motion: a non-degenerate phonon mode at a generic
wavevector can have a complex eigenvector and therefore an elliptical or circular atomic
trajectory. Conversely, when two linearly polarized modes are degenerate, circularly polarized
linear combinations can be constructed.
The mechanical angular momentum associated with the ionic motion is
\begin{equation}
\mathbf{J}_{\mathrm{ph}}
=\sum_{l\alpha}
m_\alpha\,
\mathbf{u}_{l\alpha}\times
\dot{\mathbf{u}}_{l\alpha},
\end{equation}
where \(m_\alpha\) and \(\mathbf{u}_{l\alpha}\) are the mass and displacement of atom \(\alpha\)
in unit cell \(l\). For a phonon mode \((\mathbf{k},\nu)\), we write the displacement as
\begin{equation}
\mathbf{u}_{l\alpha}(t)
=
\frac{Q_0}{\sqrt{Nm_\alpha}}
\mathrm{Re}
\left[
\enu
e^{i(\mathbf{k}\cdot\mathbf{R}_l-\omega_{\mathbf{k}\nu}t)}
\right],
\qquad
\sum_\alpha
|\enu|^2=1 ,
\end{equation}
where $N$ is the number of unit cells.
Writing the \(z\) component of the angular momentum for each atom as \(m_\alpha(u_\alpha^x\dot u_\alpha^y-u_\alpha^y\dot u_\alpha^x)\), the explicit mass dependence cancels between the prefactor and the mass normalization of the phonon
eigenvector. After averaging over one oscillation period,
\begin{equation}
J_z(\mathbf{k},\nu)
=
Q_0^2\omega_{\mathbf{k}\nu}
\sum_\alpha
\mathrm{Im}
\left[
\enux^\ast
\enuy
\right].
\end{equation}
Upon quantization, the mode-resolved angular momentum per phonon is
\begin{equation}
J_z(\mathbf{k},\nu)
=
2\hbar
\sum_\alpha
\mathrm{Im}
\left[
\enux^\ast
\enuy
\right].
\label{eq:Jz}
\end{equation}
Following Ref. \cite{Zhang_2015}, we introduce for each atom the right- and left-circular polarization vectors
\begin{equation}
|R\rangle=
\frac{1}{\sqrt{2}}(1,i)^T,
\qquad
|L\rangle=\frac{1}{\sqrt{2}}(1,-i)^T,
\end{equation}
and the corresponding projections
\begin{equation}
\begin{split}
\epsnuR\equiv\left<R\right|\left.\enu\right>
=
\frac{\enux-i \enuy}{\sqrt{2}},
\\
\epsnuL\equiv\left<L\right|\left.\enu\right>
=
\frac{\enux+i \enuy}{\sqrt{2}}.
\end{split}
\end{equation}
Equation \ref{eq:Jz} then becomes
\begin{equation}
J_z(\mathbf{k},\nu)
=
\hbar
\sum_\alpha
\left(
|\epsnuR|^2-
|\epsnuL|^2
\right).
\label{eq:Jz2}
\end{equation}
This quantity is the phonon circular polarization \(s_{\mathrm{ph}}^z\) introduced in Ref. \cite{Zhang_2015} and has the same form as the mode-resolved phonon angular momentum \(j_{\mathbf{k}\nu}^z\) of Ref. \cite{Zhang_2014}. With normalized eigenvectors, $-1\leq \displaystyle{\frac{J_z}{\hbar}}\leq 1$. Thus \(J_z/\hbar=+1\) and \(-1\) correspond to fully right- and left-circularly polarized modes, respectively, whereas intermediate values indicate a partial net circular polarization, arising from elliptical ionic motion and/or partial cancellation among the contributions of different atoms. A vanishing \(J_z\) indicates zero net circular polarization, although individual atoms may still
undergo circular motion whose contributions cancel. Although the explicit atomic masses cancel from Eq. \ref{eq:Jz2}, the phonon eigenvectors, and therefore \(J_z\), still depend on the atomic masses and force constants through the dynamical matrix.
It is important to distinguish the phonon angular momentum \(J_z\) used in this work from the phonon pseudo-angular momentum (PAM) introduced in Ref. \cite{Zhang_2015}. At a wavevector left invariant by a threefold rotation modulo a reciprocal-lattice vector, such as \(\Gamma\), \(K\), or \(K'\) in a hexagonal lattice, the phonon Bloch state can be classified ed according to
\begin{equation}
\hat C_3 \Phi_{\mathbf{k}\nu}
=
e^{-i2\pi l_{\mathrm{ph}}/3}
\Phi_{\mathbf{k}\nu},
\end{equation}
where \(l_{\mathrm{ph}}=0,\pm1\) modulo 3 is the phonon PAM. This is a symmetry quantum number and is distinct from the circular polarization \(J_z/\hbar\). Following Ref. \cite{Zhang_2015}, the PAM can be decomposed into a local or “spin” contribution \(l^s\), associated with the transformation of the intracell displacement under rotation, and an “orbital” contribution \(l^o\), originating from the Bloch phase of the intercell part. The right- and left-circular basis states are eigenstates of the local rotation with \(l^s=\pm1\), which explains why the same R/L basis naturally enters both the circular-polarization analysis and the local contribution to the PAM. However, the two quantities are not identical: the circular-polarization weight can take any value between \(-1\) and \(+1\), whereas \(l^s\) and the total \(l_{\mathrm{ph}}\) are discrete symmetry labels at the \(C_3\)-invariant wavevectors.
In the present work we use \(J_z/\hbar\), rather than \(l_{\mathrm{ph}}\), to characterize the chirality of the phonon modes. Unlike PAM, \(J_z(\mathbf{k},\nu)\) is defined throughout the Brillouin zone and can vary continuously with \(\mathbf{k}\). The \(C_3\) symmetry therefore does not, in general, quantize \(J_z/\hbar\) to \(\pm1\); these limiting values simply correspond to fully circularly polarized ionic motion. Accordingly, the color scale used in the phonon dispersions throughout this work represents \(J_z/\hbar\), i.e. the normalized phonon angular momentum or circular polarization, and not the phonon PAM \(l_{\mathrm{ph}}\).

\subsection{Results} 
\begin{figure}[h]
    \centering
    \includegraphics[scale=0.6]{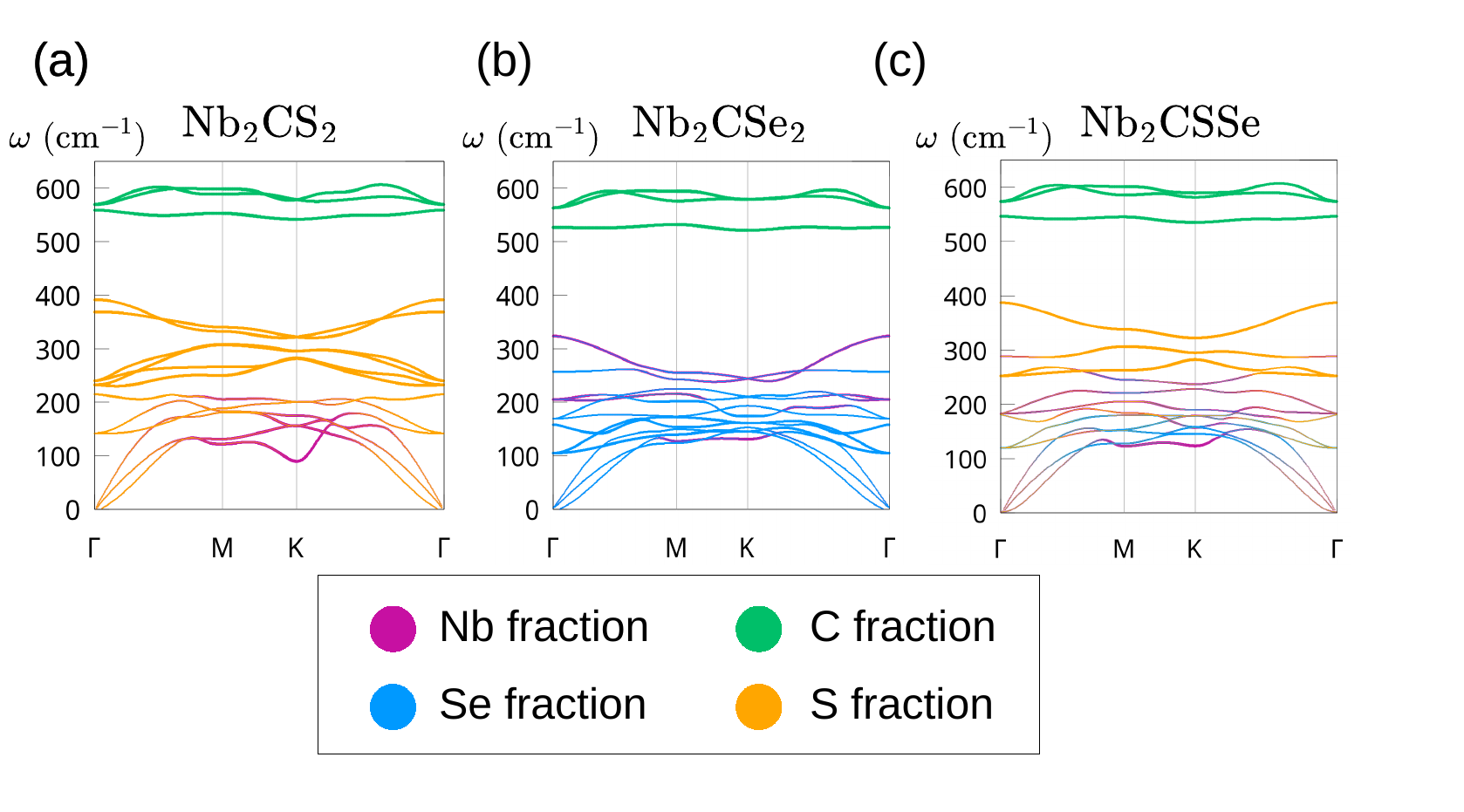}
    \caption{Phonon dispersion projected on the constituent atomic species for three representative monolayers: two TMCCs, \ce{Nb2CS2} and \ce{Nb2CSe2}, and the corresponding JTMCCs, \ce{Nb2CSSe}}
    \label{fig:TMCCphononsatoms}
\end{figure}

\begin{figure}
    \centering
    \includegraphics[scale=0.7]{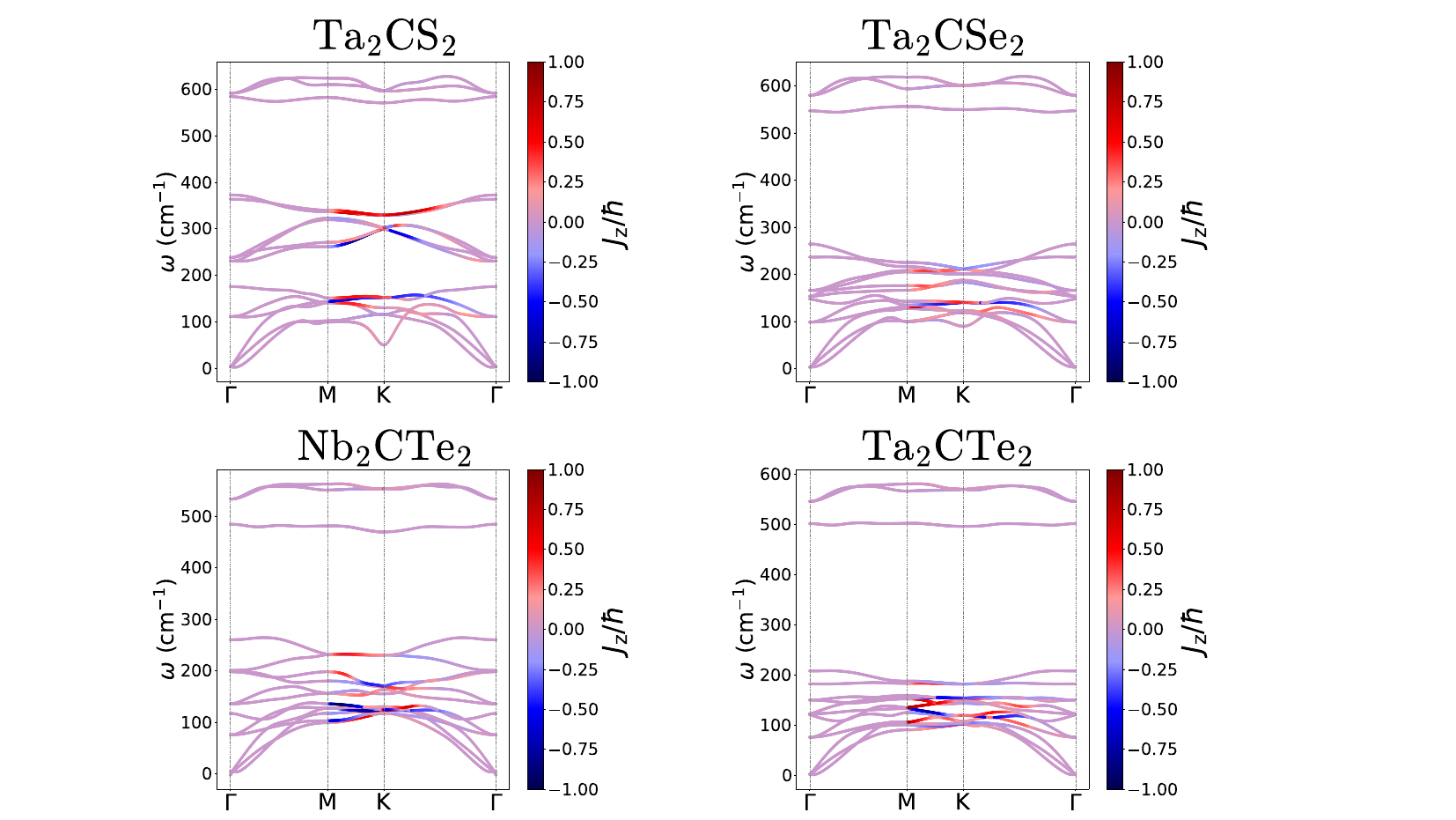}
    \caption{Phonon dispersion for TMCC monolayers which are not reported in the main text. The dispersions are colored according to the normalized phonon angular momentum $J_z/\hbar$. Blue (red) corresponds to negative (positive) circular polarization.}
    \label{fig:tmcc_ph}
\end{figure}

\begin{figure}
    \centering
    \includegraphics[scale=0.65]{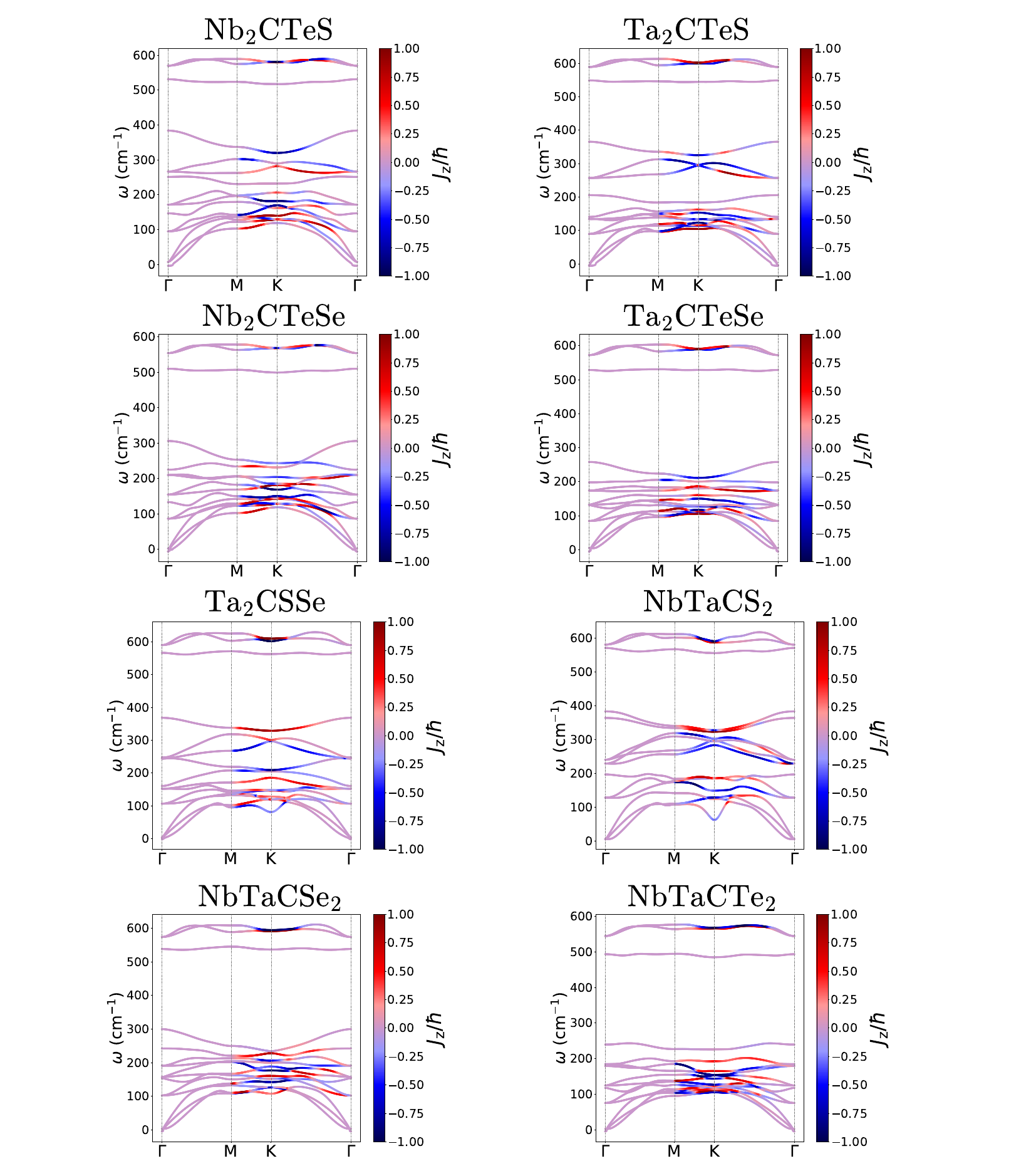}
    
    \caption{Phonon dispersion for JTMCC monolayers which are not reported in the main text. The dispersions are colored according to the normalized phonon angular momentum $J_z/\hbar$. Blue (red) corresponds to negative (positive) circular polarization. }
    \label{fig:tmccj_ph}
\end{figure}

In Fig. \ref{fig:TMCCphononsatoms} we report the phonon dispersions projected on the atomic species for three representative compounds: two TMCCs, \ce{Nb2CS2} and \ce{Nb2CSe2}, and a JTMCC, \ce{Nb2CSSe}.
It is clear that, in all the three cases, the higher frequencies are dominated by the lightest atom, that is the carbon. The others, depending on the atomic masses of the involved compounds, are more or less hybridized among the other elements. We do not report the results for the other systems, as they would be analogous.\\
In Fig. \ref{fig:tmcc_ph} we show the phonon dispersions for the TMCCs monolayers not reported in the main text. All of them result to be stable without charge-density waves (CDWs).
Projecting them on the phonon angular momentum we can also see that there is not a net and strong chirality at the $K-$point for the phonons in the last branches. This is not true for JTMCC monolayers, as illustrated in Fig. \ref{fig:tmccj_ph}.
In Tables \ref{tmcc_splitting} and \ref{tmccj_splitting}, we collect the splittings at the $K$-point for the last two dispersion bands. In the first one there are not data for $\mathcal{E}=$0, as in TMCC monolayers splittings open only with the application of an electric field.
\begin{table}[]
    \centering
\begin{tabular}{ |p{3cm}||p{3cm}|p{3cm}|  }
 \hline
 \multicolumn{3}{|c|}{TMCCs} \\
 \hline
   & $\mathcal{E}=$0.4 V/\AA & $\mathcal{E}=$-0.4 V/\AA \\
 \hline
 \ce{Nb2CS2}   & 3.17 & 3.18  \\
 \ce{Nb2CSe2}&  1.45    & 1.46  \\
 \ce{Nb2CTe2} & 0.53 & 0.53 \\
 \ce{Ta2CS2}  & 3.65 & 3.66\\
 \ce{Ta2CSe2}&   1.57  & 1.57\\
 \ce{Ta2CTe2}& 0.09  & 0.09  \\
 \hline
\end{tabular}
\caption{Splittings at the $K$-point between the last two phonon branches for TMCC monolayers in the presence of an external electric field. Frequencies are expressed in $\mathrm{cm^{-1}}$.}
\label{tmcc_splitting}
\end{table}
By symmetry, the effect is the same when changing the orientation of the out-of-plane field. The effect is more relevant in the two systems that contain the Sulfur atoms, while it is almost unnoticeable in the ones with Tellurium.\\ 
In JTMCC monolayers, splittings are induced by the built-in electric field, with a relevant amplitude in \ce{Nb2CSSe} and \ce{Ta2CSSe}.
\begin{table}[]
    \centering
\begin{tabular}{ |p{3cm}||p{3cm}|p{3cm}|p{3cm}|  }
\hline
 \multicolumn{4}{|c|}{JTMCCs} \\
 \hline
    & $\mathcal{E}=$0 & $\mathcal{E}=$0.4 V/\AA & $\mathcal{E}=$-0.4 V/\AA \\
 \hline
 \ce{Nb2CSSe}   & 7.91    & 6.31& 10.22\\
 \ce{Nb2CTeS}&   1.63 & 2.89 & 1.07  \\
 \ce{Nb2CTeSe} & 0.39 & 1.32 & 0.26 \\
 \ce{Ta2CSSe}    & 7.89 & 6.08 & 10.34 \\
 \ce{Ta2CTeS}&  1.74  & 0.65 & 2.34 \\
 \ce{Ta2CTeSe}& 1.39  & 0.67 & 1.90\\
 \ce{NbTaCS2}& 2.97  & 6.16 & 0.61\\
 \ce{NbTaCSe2}& 3.30  & 4.76 & 1.79 \\
 \ce{NbTaCTe2}&  1.91 & 2.26 & 1.56\\
 \hline
\end{tabular}
\caption{Splittings at the $K$-point between the last two phonon branches for JTMCC monolayers, without and in the presence of an external electric field. All frequencies are expressed in $\mathrm{cm^{-1}}$. For completeness, we report the exact computed values of the splitting, however the variations within 1 cm$_{-1}$ are in the DFPT error bar.}
\label{tmccj_splitting}
\end{table}
Consequenlty, these splittings can be tuned by  an external out-of-plane electric field. Their response depends on the field orientation: reversing the field can either increase or decrease the splitting, depending on its direction relative to the intrinsic Janus asymmetry.
The bigger splittings are obtained, as in the case of the external field, for the \ce{Nb2CSSe} and \ce{Ta2CSSe} monolayers, reaching values of $\sim 10\ \mathrm{cm^{-1}}$. The greater variations are, instead, obtained for the \ce{NbTaCS2} and \ce{NbTaCSe2} monolayers, in which splittings reach, respectively, approximately 6 and 5 $\mathrm{cm^{-1}}$. In all the systems (TMCCs and JTMCC) the presence of Tellurium appears to hinder such splittings and its tunability.

\section{Work functions}
Tables \ref{tmcc_work} and \ref{tmccj_work} report the work functions computed for all the compounds, at zero external electric field.
\begin{table}[h]
    \centering
\begin{tabular}{ |p{2.2cm}||p{2.2cm}| }
 \hline
 \multicolumn{2}{|c|}{TMCCs, work functions} \\
 \hline
  \ce{Nb2CS2}   & 5.60  \\
 \hline
 \ce{Nb2CSe2}  &   5.06  \\
 \hline
 \ce{Nb2CTe2} & 4.42  \\
 \hline
 \ce{Ta2CS2}  & 5.41 \\
 \hline
 \ce{Ta2CSe2}&   4.85 \\
 \hline
 \ce{Ta2CTe2}&   4.25 \\
 \hline
\end{tabular}
\caption{Work functions of the TMCC monolayers, expressed in eV.}
\label{tmcc_work}
\end{table}
\begin{table}[h]
    \centering
\begin{tabular}{ |p{2.6cm}||p{2.6cm}|p{2.6cm}| }
 \hline
 \multicolumn{3}{|c|}{JTMCCs, work functions} \\
 \hline
  \ce{Nb2CSSe} & 5.64 (S-side)& 4.97 (Se-side) \\
 \hline
 \ce{Nb2CTeS}  &  5.77 (S-side)  & 4.28 (Te-side) \\
 \hline
 \ce{Nb2CTeSe} & 5.21 (Se-side)  & 4.35 (Te-side) \\
 \hline
 \ce{Ta2CSSe}  & 5.43 (S-side) & 4.78 (Se-side) \\
 \hline
 \ce{Ta2CTeS}&  5.53 (S-side) & 4.05 (Te-side) \\
 \hline
 \ce{Ta2CTeSe}& 4.98 (Se-side)   & 4.12 (Te-side)\\
 \hline
 \ce{NbTaCS2}&  5.62 (Nb-side)  & 5.41 (Ta-side) \\
 \hline
 \ce{NbTaCSe2}& 5.04 (Nb-side)  & 4.88 (Ta-side) \\
 \hline
 \ce{NbTaCTe2}& 4.40  (Nb-side) & 4.28 (Ta-side)\\
 \hline
\end{tabular}
\caption{Work functions computed at the two terminations of JTMCC monolayers. Values are expressed in eV. }
\label{tmccj_work}
\end{table}
For TMCC, by symmetry, we have only one value of the work function, while in Janus monolayers two distinct values are found at the two different sides of the slab.

  \printbibliography[title={Supporting References}]

@article{Kumbhakar_2023,
    author  = {Kumbhakar, Partha and Jayan, Jitha S. and Madhavikutty, Athira Sreedevi and others},
    title   = {Prospective applications of two-dimensional materials beyond laboratory frontiers: a review},
    journal = {Science},
    volume  = {26},
    number  = {5},
    pages   = {106671},
    year    = {2023},
    issn    = {2589-0042},
}

@article{Sun_2026,
title = {The rise of chiral phononics: From fundamental concepts to emergent quasiparticle interactions},
journal = {Physics Reports},
volume = {1195},
pages = {1-99},
year = {2026},
issn = {0370-1573},
author = {Hong Sun and Tingting Wang and Zihan Chen and Yi Wang and Xiaozhe Li and Hao Chen and Lifa Zhang},
}

@article{Wolf2001,
  author = {Wolf, S. A. and Awschalom, D.D. and Buhrman, R. A. and others},
  title = {Spintronics: a spin-based electronics vision for the future},
  journal = {Science},
  year = {2001},
  volume = {294},
  number = {5546},
  pages = {1488--1495},
  doi = {10.1126/science.1065389}
}

@article{Ma2022,
  author = {Ma, Q. and Yang, G. and Liu, Y.},
  title = {Two-dimensional ferromagnetic half-metallic {J}anus {V}\textsubscript{2}{A}s{P} monolayer},
  journal = {Applied Physics Letters},
  year = {2022},
  volume = {120},
  pages = {123105},
  doi = {10.1063/5.0090654}
}

@article{Zhang2022,
  author = {Zhang, L. and Xia, Y. and Li, X. and others},
  title = {Janus transition-metal dichalcogenides},
  journal = {Journal of Applied Physics},
  year = {2022},
  volume = {131},
  pages = {230902},
  doi = {10.1063/5.0086432}
}

@article{Hamada2018,
  author = {Hamada, M. and Minamitani, E. and Hirayama, M. and Murakami, S.},
  title = {Phonon angular momentum induced by the temperature gradient},
  journal = {Physical Review Letters},
  year = {2018},
  volume = {121},
  pages = {175301},
  doi = {10.1103/PhysRevLett.121.175301}
}

@article{Shen_2017,
  title = {Novel Mechanism of a Charge Density Wave in a Transition Metal Dichalcogenide},
  author = {Shen, D. W. and Xie, B. P. and Zhao, J. F. and others},
  journal = {Phys. Rev. Lett.},
  volume = {99},
  issue = {21},
  pages = {216404},
  numpages = {4},
  year = {2007},
  month = {Nov},
  publisher = {American Physical Society},
}

@article{Zhang_2021,
  title = {Visualization of edge-modulated charge-density-wave orders in monolayer transition-metal-dichalcogenide metal},
  author = {Zhang, Q. and Fan, J. and Zhang, T. and others},
  journal = {Commun. Phys.},
  volume = {5},
  pages = {117},
  year = {2022},
}

@article{Tang_2022,
author = {Tang, Xiao and Kou, Liangzhi},
title = {2{D} Janus Transition Metal Dichalcogenides: Properties and Applications},
journal = {Physica Status Solidi (B)},
volume = {259},
number = {4},
pages= {2100562},
year = {2022}
}

@Article{Pang_2020,
author ="Pang, Xin and Wu, Tong and Gu, Yuhao and others",
title  ="\ce{{N}b2{S}e2{C}}: a new compound as a combination of transition metal dichalcogenide and {MXene} for oxygen evolution reaction",
journal  ="Chem. Commun.",
year  ="2020",
volume  ="56",
issue  ="63",
pages  ="9036-9039",
}

@article{Bordoloi_2024,
    author = {Bordoloi, Arjyama and Garcia-Castro, A. C. and Romestan, Zachary and others},
    title = {Promises and technological prospects of two-dimensional {R}ashba materials},
    journal = {Journal of Applied Physics},
    volume = {135},
    number = {22},
    pages = {220901},
    year = {2024},
    month = {06},
    issn = {0021-8979},
}

@article{Bihlmayer_2022,
    author  = {Bihlmayer, G. and No{\"e}l, P. and Vyalikh, Denis V. and others},
    title   = {Rashba-like physics in condensed matter},
    journal = {Nat. Rev. Phys.},
    volume  = {4},
    number  = {10},
    pages   = {642--659},
    year    = {2022},
    month   = {08},
}

@article{Manchon_2015,
    author = {Manchon, A. and Koo, H. and Nitta, J. and others},
    title = {New perspectives for {R}ashba spin–orbit coupling},
    journal = { Nature Mater},
    volume = {14},
    pages = {871-882},
    year = {2015},
}

@article{Kassa_2024,
title = {Emerging 2{D} materials beyond {MX}enes and {T}{M}{D}s: Transition metal carbo-chalcogenides},
journal = {Progress in Materials Science},
volume = {144},
pages = {101287},
year = {2024},
issn = {0079-6425},
author = {Ibrahim, K. B and Shifa, T. A. and Zorzi, S. and others},
}

@article{Unnati_2024,
title = {Unveiling the potential of newly synthesized 2{D} {T}{M}{C}{C} monolayer for {H}2{S} gas sensing: A {D}{F}{T} study},
journal = {Surfaces and Interfaces},
volume = {54},
pages = {105098},
year = {2024},
issn = {2468-0230},
author = {Jethawa, U. and Chaudhari, A and Chakraborty, B.},
}

@article{Majed_2022,
author = {Majed, Ahmad and Kothakonda, Manish and Wang, Fei and others},
title = {Transition Metal Carbo-Chalcogenide “{T}{M}{C}{C}:” A New Family of 2{D} Materials},
journal = {Advanced Materials},
volume = {34},
number = {26},
pages = {2200574},
year = {2022}
}

@article{Ren_2025,
author = {Ren, K. and Pan, Z. and Tu, X. and others},
title = {2{D} Transition Metal Carbo-Chalcogenide \ce{{N}b2{C}{S}e2} : A Combination of Transition Metal Dichalcogenide and {MX}ene for Ultrafast Photonics Application},
journal = {Advanced Optical Materials},
volume = {13},
pages = {2402385},
year = {2025}
}

@article{Jing_2019,
author = {Jing, Yu and Liu, Jie and Zhou, Zhenpei and others},
title = {Metallic \ce{{N}b2{S}2{C}} Monolayer: A Promising Two-Dimensional Anode Material for Metal-Ion Batteries},
journal = {The Journal of Physical Chemistry C},
volume = {123},
number = {44},
pages = {26803-26811},
year = {2019},
}

@article{Lakshmy_2024,
author = {Lakshmy, Seetha and Kalarikkal, Nandakumar and Chakraborty, Brahmananda},
title = {Transition Metal ({C}u, {P}d, and {A}g)-Modified \ce{{N}b2{S}2{C}} Monolayer for Highly Efficient Catechol Sensing: A First-Principles Investigation},
journal = {Langmuir},
volume = {40},
number = {27},
pages = {13819-13833},
year = {2024},
}

@article{Helmer_2024,
author = {Helmer, Pernilla and Bj{\"o}rk, Jonas and Rosen, Johanna},
title = {Computational Screening of Chalcogen-Terminated Inherent Multilayer MXenes and {M}2{A}{X} Precursors},
journal = {Inorganic Chemistry},
volume = {63},
number = {36},
pages = {16645-16654},
year = {2024},
}

@article{Wang_2025,
author = {Wang, J. and Cheng, R. and Hu, T. and others},
title = {Layered metal carbo–selenide \ce{{N}b2{C}{S}e2} with van der {W}aals interlayer coupling},
year = {2025},
journal = {Journal of Advanced Ceramics},
volume = {14},
number = {1},
pages = {9221008},
}

@article{Juraschek_2025,
    author = "Juraschek, D.M. and Geilhufe, R.M. and Zhu, H. and others",
    title = "Chiral phonons",
    journal =  "Nat. Phys.",
    year = "2025",
    volume="21",
    pages="1532–1540",
}

@article{Schwerdtfeger_2011,
author = {Schwerdtfeger, Peter},
title = {The Pseudopotential Approximation in Electronic Structure Theory},
journal = {ChemPhysChem},
volume = {12},
number = {17},
pages = {3143-3155},
year = {2011}
}

@article{Bengtsson_1999,
  title = {Dipole correction for surface supercell calculations},
  author = {Bengtsson, Lennart},
  journal = {Phys. Rev. B},
  volume = {59},
  issue = {19},
  pages = {12301--12304},
  year = {1999},
  month = {May},
  publisher = {American Physical Society},
}

@article{Giannozzi_2009,
year = {2009},
month = {sep},
volume = {21},
number = {39},
pages = {395502},
author = {Giannozzi, Paolo and Baroni, Stefano and Bonini, Nicola and others},
title = {QUANTUM {ESPRESSO}: a modular and open-source software project for quantum
simulations of materials},
journal = {Journal of Physics: Condensed Matter},
}

@article{Giannozzi_2017,
year = {2017},
month = {oct},
publisher = {IOP Publishing},
volume = {29},
number = {46},
pages = {465901},
author = {Giannozzi, P. and Andreussi, O. and Brumme, T. and others},
title = {Advanced capabilities for materials modelling with {Q}uantum {ESPRESSO}},
journal = {Journal of Physics: Condensed Matter},
}

@article{Giannozzi_2020,
    author = {Giannozzi, Paolo and Baseggio, Oscar and Bonfà, Pietro and others},
    title = {Quantum {ESPRESSO} toward the exascale},
    journal = {The Journal of Chemical Physics},
    volume = {152},
    number = {15},
    pages = {154105},
    year = {2020},
    month = {04},
    issn = {0021-9606},
}

@article{Blochl_1994,
  title = {Projector augmented-wave method},
  author = {Bl\"ochl, P. E.},
  journal = {Phys. Rev. B},
  volume = {50},
  issue = {24},
  pages = {17953--17979},
  numpages = {0},
  year = {1994},
  month = {Dec},
  publisher = {American Physical Society},
}

@article{Rivera_2026,
title = {Janus transition-metal carbo-chalcogenide \ce{{N}b2{C}{S}{S}e} anode enabling efficient {L}i storage},
journal = {Journal of Energy Storage},
volume = {151},
pages = {120380},
year = {2026},
issn = {2352-152X},
author = { Rivera, V. J. R. and Gonzalo, F. M. and  others},
}

@article{Rivera_2_2026,
title = {\ce{NbTaCSe2} {J}anus carbo-chalcogenide as a high-performance anode for lithium-ion batteries},
author = {Rivera, V. J. R. and Gonzalo, F. M. and others},
journal = {Applied Surface Science},
volume = {747},
pages = {167694},
year = {2026},
issn = {0169-4332},
doi = {https://doi.org/10.1016/j.apsusc.2026.167694}
}

@article{Hu_2018,
  title = {Intrinsic and anisotropic Rashba spin splitting in {J}anus transition-metal dichalcogenide monolayers},
  author = {Hu, Tao and Jia, Fanhao and Zhao, Guodong and others},
  journal = {Phys. Rev. B},
  volume = {97},
  issue = {23},
  pages = {235404},
  numpages = {6},
  year = {2018},
  month = {Jun},
}

@article{Sufyan2026,
  author = {Sufyan, Ali and Larsson, J. Andreas and Kreisel, Andreas and van Loon, Erik},
  title = {Giant {R}ashba splitting and enhanced nonlinear {B}erry-phase responses in sliding-tunable vd{W} {MXene} heterostructures},
  eprint = {2605.09674},
  journal = {arXiv},
  primaryClass = {cond-mat.mtrl-sci},
  year = {2026},
  month = {May},
  doi = {10.48550/arXiv.2605.09674}
}

@article{Zhang_2015,
  title = {Chiral Phonons at High-Symmetry Points in Monolayer Hexagonal Lattices},
  author = {Zhang, Lifa and Niu, Qian},
  journal = {Phys. Rev. Lett.},
  volume = {115},
  issue = {11},
  pages = {115502},
  numpages = {5},
  year = {2015},
  month = {Sep},
  publisher = {American Physical Society},
}

@article{Sahu_2026,
  title = {Chiral orbital and spin textures, and {E}delstein effects in monolayer {J}anus transition metal dichalcogenides},
  author = {Sahu, Pratik and Satpathy, S. and Nanda, B. R. K.},
  journal = {Phys. Rev. B},
  volume = {113},
  issue = {13},
  pages = {134435},
  numpages = {10},
  year = {2026},
  month = {Apr},
  publisher = {American Physical Society},
}

@Article{Szalowski_2024,
AUTHOR = {Szałowski, Karol},
TITLE = {Janus Monolayer of {1T-\ce{TaSSe}}: A Computational Study},
JOURNAL = {Materials},
VOLUME = {17},
YEAR = {2024},
NUMBER = {18},
PAGES = {4591},
ISSN = {1996-1944},
}

@Article{Vu_2026,
author ="Vu, Tuan V. and Kartamyshev, A. I. and Hoi, Bui D. and others",
title  ="Rashba spin splitting and phonon-limited carrier mobility in novel two-dimensional Janus \ce{{Z}{C}r{B}{S}e2} {(Z = N{,} P{,} As)} materials",
journal  ="RSC Adv.",
year  ="2026",
volume  ="16",
issue  ="17",
pages  ="15441-15451",
publisher  ="The Royal Society of Chemistry",
}

@article{Xiaojuan_2022,
author = {Ni, X. and Li, H. and Brédas, J. L.},
title = {Organic self-assembled monolayers on superconducting \ce{{N}b{S}e2}: interfacial electronic structure and energetics},
journal = {J. Phys. Condens. Matter},
volume = {34},
pages = {294003},
number = {29},
year = {2022},
}

@misc{Barreca,
    author = {Barreca, Valentina and Sangiovanni, Giorgio and Cantele, Giovanni and Zullo, Ludovica},
    title={{J}anus {A}symmetry {U}nlocks {R}ashba {S}pintronics and {C}hiral {P}hononics in {T}wo-{D}imensional {C}arbo-{C}halcogenides - {S}upplementary {I}nformation},
}

@article{Feng_2024,
author = {Feng, Runfa and Wang, Wei and Bao, C. and others},
title = {Selective Control of Phases and Electronic Structures of Monolayer \ce{{T}a{T}e2}},
journal = {Advanced Functional Materials},
volume = {36},
issue= {3},
year = {2024},
pages={2302297}
}

@article{Barajas_2018,
author = {Barajas-Aguilar, A.H. and Irwin, J.C. and Garay-Tapia, A.M. and others},
title = {Crystalline structure, electronic and lattice-dynamics properties of \ce{{N}b{T}e2}},
journal = {Sci Rep},
volume = {8},
year = {2018},
pages = {16984}
}

@misc{qeplotter,
  author       = {Shubics},
  title        = {{QEPlotter}: Quantum {ESPRESSO} band structure, {DOS}/{PDOS}, and fatband plotting toolkit},
  year         = {2024},
  publisher    = {GitHub},
  journal      = {GitHub repository},
  howpublished = {\url{https://github.com/shubics/QEPlotter}}
}

@article{kokalj_2003,
  author    = {Kokalj, Anton},
  title     = {Computer graphics and graphical user interfaces as tools in simulations of matter at the atomic scale},
  journal   = {Computational Materials Science},
  volume    = {28},
  number    = {2},
  pages     = {155--168},
  year      = {2003},
  publisher = {Elsevier},
}

@article{Lin_2020,
author = {Lin, D. and Li, S and Wen, J. and others},
title = {Patterns and driving forces of dimensionality-dependent charge density waves in 2{H}-type transition metal dichalcogenides},
journal = {Nature Communications},
volume = {11},
year = {2020},
}

@article{Zullo_2023,
author = {Zullo, Ludovica and Marini, Giovanni and Cren, Tristan and Calandra, Matteo},
title = {Misfit Layer Compounds as Ultratunable Field Effect Transistors: From Charge Transfer Control to Emergent Superconductivity},
journal = {Nano Letters},
volume = {23},
number = {14},
pages = {6658-6663},
year = {2023},
}

@article{Wang_2026,
author = {Wang, Yu and Li, Mian and Ouyang, Xiaoping and others},
title = {2{D} Transition Metal Carbo-Chalcogenides: An Emerging Family of Layered Materials},
journal = {ChemPhysChem},
volume = {27},
number = {15},
pages = {e202500895},
year = {2026}
}

@article{Yao_2017,
  title = {Manipulation of the large {R}ashba spin splitting in polar two-dimensional transition-metal dichalcogenides},
  author = {Yao, Qun-Fang and Cai, Jia and Tong, Wen-Yi and others},
  journal = {Phys. Rev. B},
  volume = {95},
  issue = {16},
  pages = {165401},
  numpages = {8},
  year = {2017},
  month = {Apr},
  publisher = {American Physical Society},
}

@article{Wick_2015,
    author = {Wickramaratne, Darshana and Zahid, Ferdows and Lake, Roger K.},
    title = {Electronic and thermoelectric properties of van der {W}aals materials with ring-shaped valence bands},
    journal = {Journal of Applied Physics},
    volume = {118},
    number = {7},
    pages = {075101},
    year = {2015},
    month = {08},
}

@article{Zhu_2018,
author = {Hanyu, Zhu  and Jun, Yi  and Ming-Yang, Li  and others },
title = {Observation of chiral phonons},
journal = {Science},
volume = {359},
number = {6375},
pages = {579-582},
year = {2018},
}

@article{Wang2026,
    author = {Wang, Hanyu and Li, Haoshu and Chen, Hao and Zhang, Lifa},
    title = {Phonon chirality and {H}all effect in {J}anus materials},
    journal = {Applied Physics Letters},
    volume = {128},
    number = {11},
    pages = {112202},
    year = {2026},
    month = {03},
    issn = {0003-6951},
    doi = {10.1063/5.0319243}
}

@article{Zullo_2024_PhysRevB.110.165411,
  title = {Twist-tunable spin control in twisted bilayer bismuthene},
  author = {Zullo, Ludovica and Ninno, Domenico and Cantele, Giovanni},
  journal = {Phys. Rev. B},
  volume = {110},
  issue = {16},
  pages = {165411},
  numpages = {9},
  year = {2024},
  month = {Oct},
  publisher = {American Physical Society},
  doi = {10.1103/PhysRevB.110.165411},
}

@article{Ernzerhof_1999,
    author = {Ernzerhof, Matthias and Scuseria, Gustavo E.},
    title = "{Assessment of the Perdew–Burke–Ernzerhof exchange-correlation functional}",
    journal = {The Journal of Chemical Physics},
    volume = {110},
    number = {11},
    pages = {5029-5036},
    year = {1999},
}

@article{DalCorso_2014,
title = {Pseudopotentials periodic table: From H to Pu},
journal = {Computational Materials Science},
volume = {95},
pages = {337-350},
year = {2014},
author = {Dal Corso, Andrea},
}

@article{Zhang_2014,
  title = {Angular Momentum of Phonons and the Einstein--de Haas Effect},
  author = {Zhang, Lifa and Niu, Qian},
  journal = {Phys. Rev. Lett.},
  volume = {112},
  issue = {8},
  pages = {085503},
  numpages = {5},
  year = {2014},
  month = {Feb},
  publisher = {American Physical Society},
  doi = {10.1103/PhysRevLett.112.085503},
  url = {https://link.aps.org/doi/10.1103/PhysRevLett.112.085503}
}
\end{refsection}

\end{document}